\documentclass[12pt,preprint]{aastex}

\newcommand{\msunyr}{M$_{\odot}$~yr$^{-1}$}

\shorttitle{Comparison of UV and H$\alpha$ SFRs}
\shortauthors{Lee et al.}

\begin{document}

\title{Comparison of H$\alpha$ and UV Star Formation Rates \\ in the Local Volume: \\ Systematic Discrepancies for Dwarf Galaxies} 

\author{
Janice C. Lee\altaffilmark{1,2}, 
Armando Gil de Paz\altaffilmark{3},
Christy Tremonti\altaffilmark{4,15,16},
Robert C. Kennicutt, Jr.\altaffilmark{5,6},
Samir Salim\altaffilmark{7},
Matthew Bothwell\altaffilmark{5}
Daniela Calzetti\altaffilmark{8},
Julianne Dalcanton\altaffilmark{9},
Daniel Dale\altaffilmark{10},
Chad Engelbracht\altaffilmark{5},
Jos\'e G. Funes, S.J.\altaffilmark{11},
Benjamin Johnson\altaffilmark{5},
Shoko Sakai\altaffilmark{12},
Evan Skillman\altaffilmark{13},
Liese van Zee\altaffilmark{14},
Fabian Walter\altaffilmark{4},
Daniel Weisz\altaffilmark{13}
}

\altaffiltext{1}{Carnegie Observatories, 813 Santa Barbara Street, Pasadena, CA 91101; jlee@ociw.edu}
\altaffiltext{2}{Hubble Fellow}
\altaffiltext{3}{Departmento de Astrofisica, Universidad Complutense de Madrid, Madrid 28040, Spain}
\altaffiltext{4}{Max-Planck-Institut f\"{u}r Astronomie, K\"{o}nigstuhl 17, D-69117 Heidelberg, Germany}
\altaffiltext{5}{Institute of Astronomy, University of Cambridge, Madingley Road, Cambridge CB3 0HA, UK}
\altaffiltext{6}{Steward Observatory, University of Arizona, Tucson, AZ 85721}
\altaffiltext{7}{NOAO, 950 North Cherry Ave., Tucson, AZ 85719}
\altaffiltext{8}{Department of Astronomy, University of Massachusetts, Amherst, MA 01003}
\altaffiltext{9}{Department of Astronomy, University of Washington, Box 351580, Seattle, WA 98195}
\altaffiltext{10}{Department of Physics \& Astronomy, University of Wyoming, Laramie, WY 82071} 
\altaffiltext{11}{Vatican Observatory, Specola Vaticana, V00120 Vatican}
\altaffiltext{12}{Division of Astronomy and Astrophysics, University of California, Los Angeles, Los Angeles, CA, 90095-1562}
\altaffiltext{13}{Department of Astronomy, University of Minnesota, Minneapolis, MN 55455} 
\altaffiltext{14}{Astronomy Department, Indiana University, Bloomington, IN 47405}
\altaffiltext{15}{Alexander von Humbolt Fellow}
\altaffiltext{16}{Department of Astronomy, University of Wisconsin-Madison, 475 N. Charter Street, Madison WI 53706-1582}

\begin{abstract}

Using a complete sample of $\sim$300 star-forming galaxies within 11 Mpc of the Milky Way, we evaluate the consistency between star formation rates (SFRs) inferred from the far ultraviolet (FUV) non-ionizing continuum and H$\alpha$ nebular emission, assuming standard conversion recipes in which the SFR scales linearly with luminosity at a given wavelength.  Our analysis probes SFRs over 5 orders of magnitude, down to ultra-low activities on the order of $\sim$10$^{-4}$ \msunyr.  The data are drawn from the 11 Mpc H$\alpha$ and Ultraviolet Galaxy Survey (11HUGS), which has obtained H$\alpha$ fluxes from ground-based narrowband imaging, and UV fluxes from imaging with GALEX.  For normal spiral galaxies (SFR$\sim$1 \msunyr), our results are consistent with previous work which has shown that FUV SFRs tend to be lower than H$\alpha$ SFRs before accounting for internal dust attenuation, but that there is relative consistency between the two tracers after proper corrections are applied.  However, a puzzle is encountered at the faint end of the luminosity function.  As lower luminosity dwarf galaxies, roughly less active than the Small Magellanic Cloud, are examined, H$\alpha$ tends to increasingly under-predict the total SFR relative to the FUV.  The trend is evident prior to corrections for dust attenuation, which affects the FUV more than the nebular H$alpha$ emission, so this general conclusion is robust to the effects of dust.  Although past studies have suggested similar trends, this is the first time this effect is probed with a statistical sample for galaxies with SFR$\lesssim$0.1 \msunyr.  By SFR$\sim$0.003 \msunyr, the average H$\alpha$-to-FUV flux ratio is lower than expected by a factor of two, and at the lowest SFRs probed, the ratio exhibits an order of magnitude discrepancy for the handful of galaxies that remain in the sample.  
A range of standard explanations does not appear to be able to account for the magnitude of the systematic.  Some recent work has argued for an IMF which is deficient in high mass stars in dwarf and low surface brightness galaxies, and we also consider this scenario.  Under the assumption that the FUV traces the SFR in dwarf galaxies more robustly, the prescription relating H$\alpha$ luminosity to SFR is re-calibrated for use in the low SFR regime when FUV data are not available. 

\end{abstract}


\section{Introduction}
\label{sec:intro}

H$\alpha$ nebular line emission and the ultraviolet (UV) non-ionizing continuum flux are two fundamental, widely-used star formation rate (SFR) indicators.  H$\alpha$ nebular emission arises from the recombination of gas ionized by the most massive O- and early-type B-stars ($M_{\ast}\gtrsim17 M_{\odot}$). It therefore traces star formation over the lifetimes of these stars, which is on the order of a few million years.  In contrast, the UV flux primarily originates from the photospheres of a fuller mass spectrum of O- through later-type B-stars ($M_{\ast}\gtrsim3 M_{\odot}$), and thus measures star formation averaged over a longer $\sim10^8$ yr timescale.

In principle, the two tracers should yield consistent SFRs.  Discrepancies that are otherwise found can be valuable in revealing problems in the calibrations of the indicators that are used and/or the assumptions underlying them.  A number of studies have directly compared the integrated H$\alpha$ and UV SFRs of galaxies for which both diagnostics have been measured (e.g., Buat et al. 1987; Buat 1992; Sullivan et al. 2000; Bell \& Kennicutt 2001; Buat et al. 2002; Iglesias-Paramo et al. 2004; Salim et al. 2007), with most assuming standard conversion recipes where the ratios of the UV and H$\alpha$ luminosities to the SFR are constants (e.g. Kennicutt 1998; hereafter K98).  For normal spiral galaxies (SFR$\gtrsim$1 \msunyr), there is broad agreement among such studies that although the UV SFR tends to be lower than the H$\alpha$ SFR before accounting for internal dust attenuation, there is good consistency between the two after proper corrections have been applied.  This implies that the UV stellar continuum is more affected by attenuation than the H$\alpha$ nebular emission, and the intrinsic dust-corrected H$\alpha$-to-UV flux ratio is constant on average.  As lower mass dwarf galaxies are probed however, there have been some indications that the observed H$\alpha$-to-UV flux ratio may systematically decrease, and SFRs inferred from H$\alpha$  will tend to be lower relative to the UV (e.g., Sullivan et al. 2000; Bell \& Kennicutt 2001).  Such a trend could provide evidence, for example, that the recent star formation histories of low-mass galaxies are predominantly bursty, or otherwise exhibit abrupt changes.  Following a discontinuity in star formation, a dearth of ionizing stars develops as those stars expire relative to the longer-lived, lower mass stars that also significantly contribute to the UV emission, and this may cause the H$\alpha$-to-UV flux ratio to be lower than expected (e.g., Sullivan et al. 2004; Iglesias-Paramo et al. 2004).  Or it may perhaps suggest that the stellar initial mass function is not universal and becomes more deficient in the highest mass ionizing stars within low-mass, low-surface brightness galaxies (e.g., Meurer et al. 2009; Pflamm-Altenburg et al. 2009).  However, wholesale deviations from the expected ratio as a function of luminosity have only begun to become apparent when integrated SFRs less than about 0.1 \msunyr\ have been probed (i.e., a few times lower than the SFR of the Large Magellanic Cloud), and only handfuls of galaxies with such low activities were included in the previous studies.  Therefore, the observational trend first needs to be more robustly established, preferably with an unbiased, statistical sample of dwarf galaxies.  

In order to further study the relationship between the UV and H$\alpha$ emission, 
we draw upon GALEX UV observations for a complete sub-set of star-forming galaxies from the 
11 Mpc H$\alpha$ narrowband imaging survey of Kennicutt et al. (2008).  
Dwarf galaxies dominate this sample by number, and over 80\% have H$\alpha$-based SFRs$<$0.1 \msunyr.  This dataset is thus ideal for probing the relationship between the UV and H$\alpha$ emission in the low SFR regime.  Such a sample is also particularly well-suited for study in the UV since the vast majority of the galaxies have low-metallicities and low-dust contents which minimizes difficulties with otherwise vexing attenuation corrections.  

The precursor H$\alpha$ survey and GALEX follow-up program together has been referred to as 11HUGS, the 11 Mpc H$\alpha$ and Ultraviolet Galaxy Survey.  The data from 11HUGS are part of a broader campaign designed to provide a census of star formation and dust in the Local Volume.  11HUGS is being further augmented by Spitzer IRAC mid-infrared and MIPS far-infrared observations through the composite Local Volume Legacy\footnote{http://www.ast.cam.ac.uk/research/lvls/} (LVL) program (Lee et al. 2008; Dale et al. 2009), which is delivering the overall multi-wavelength dataset to the community.\footnote{Public data releases have begun through the NASA/IPAC Infrared Science Archive (IRSA; http://ssc.spitzer.caltech.edu/legacy/lvlhistory.html).}  
The ensemble of observations enables studies ranging from detailed spatially-resolved analyses probing the properties of individual HII regions and the radial profiles of disks (e.g., Munoz-Mateos et al. 2009), to a broader statistical examination of the demographics of the local star-forming galaxy population as a whole (e.g., Lee et al. 2007, Dale et al. 2009).  Other specific studies currently being pursued by the 11HUGS and LVL teams include the calibration of monochromatic far-IR SFR indicators (Calzetti et al. 2009) and photometric measurement indicies for PAH emission (Marble et al. 2009), and an examination of the prevalence and duty cycle of starbursts in dwarf galaxies (Lee et al. 2009a).

An outline of this paper is as follows.  
In Section 2, we describe the 11HUGS H$\alpha$ narrowband and GALEX UV datasets which are used in this analysis.
In Section 3,  we compute SFRs and evaluate the consistency between the two tracers over 5 orders of magnitude down to ultra-low activities of 10$^{-4}$ \msunyr.  We adopt the SFR conversion prescriptions of K98, so this is equivalent to examining whether the H$\alpha$-to-UV flux ratio exhibits systematic deviations from a constant value.  Our primary observational result is that the H$\alpha$-to-FUV flux ratio decreases with decreasing B-band, H$\alpha$ and FUV integrated luminosities.  This systematic is evident prior to any corrections for internal dust attenuation, which affects the FUV more than the H$\alpha$ nebular emission.  Our analysis provides a robust confirmation of trends suggested by prior studies, with extension to lower SFRs than previously probed.  In Section 4, we investigate possible causes of this trend.  The examination of the assumptions underlying the K98 calibration provides a convenient framework for our analysis, and we check the validity of these assumptions with respect to the local dwarf galaxy population.  We discuss the effects of uncertainties in the stellar evolution tracks and model atmospheres, non-solar metallicities, non-constant SFHs, possible leakage of ionizing photons/departures from Case B recombination, dust attenuation and stochasticity in the formation of high-mass stars.  However, none of these drivers, at least when considered individually, appear to be able to account for the observed trend.  We therefore also consider systematic variations in the IMF where dwarf galaxies with lower activities are deficient in the most massive ionizing stars.  At the end of Section 4, we re-evaluate H$\alpha$ as a SFR indicator for dwarf galaxies, and provide an alternate conversion prescription for use when FUV luminosities are not available.  We conclude in Section 5 by summarizing our results, and outline remaining uncertainties in the interpretation and possible directions for future work.

\section{Data}

To investigate the relationship between the FUV and H$\alpha$ emission in the low luminosity regime, we draw upon data collected by the 11HUGS GALEX Legacy program.  The 11HUGS sample is dominated by dwarf galaxies and is thus ideal for studying the nature of galaxies with low SFRs.  Integrated H$\alpha$ and UV flux catalogs are provided by Kennicutt et al. (2008; hereafter Paper I) and Lee et al. (2009b; hereafter Paper II), respectively.  Details on the sample selection, observations, photometry and general properties of the sample are given in those papers.  Here, we provide a brief summary of the dataset.

Our overall Local Volume sample of 436 objects is compiled from existing galaxy catalogs (as described in Paper I), and is divided into two components.  The primary component aims to be as complete as possible in its inclusion of known nearby star-forming galaxies within given limits, and consists of spirals and irregulars (T$\geq$0) within 11 Mpc that avoid the Galactic plane ($|b|>20^{\circ}$) and are brighter than $B=15$ mag.  These bounds represent the ranges within which the original surveys that have provided the bulk of our knowledge on the Local Volume galaxy population have been shown to be relatively complete (e.g., Tully 1988, deVaucouleurs et al. 1991), while still spanning a large enough volume to probe a diverse cross-section of star formation properties.  A secondary component consists of galaxies that are still within 11 Mpc, but fall outside one of the limits on brightness, Galactic latitude, or morphological type, and have available H$\alpha$ flux measurements (i.e., targets that were either observed by our group as telescope time allowed, or had H$\alpha$ fluxes published in the literature).  Subsequent statistical tests, as functions of the compiled B-band apparent magnitudes and 21-cm (atomic hydrogen, HI) flux, show that the overall sample of star-forming galaxies is complete to $\sim$15.5 mag and 6 Jy km s$^{-1}$, respectively (Lee et al. 2009a).  These limits correspond to $M_B\lesssim-15$ and $M_{HI}> 2 \times 10^8$ M$_{\odot}$ for $|b|>20^{\circ}$ at the edge of the 11 Mpc volume.  

Through a combination of new narrowband H$\alpha$+[NII] and $R$-band imaging, and data compiled from the literature, Paper I provides integrated H$\alpha$ fluxes for over 90\% of the total sample.  The new narrowband data were obtained at 1-2 m class telescopes in both hemispheres, and reached relatively deep point source flux and surface brightness limits of $\sim$2 $\times$ 10$^{-16}$ ergs cm$^{-2}$ s$^{-1}$ and $\sim$4 $\times$ 10$^{-18}$ ergs cm$^{-2}$ s$^{-1}$ arcsec$^{-2}$, respectively.  

The GALEX UV follow-up imaging primarily targeted the $|b|>30^{\circ}$, $B<15.5$ subset of galaxies.  The more restrictive latitude limit was imposed to avoid excessive Galactic extinction and fields with bright foreground stars and/or high background levels for which UV imaging would be prohibited due to the instrument's brightness safety limits.  The GALEX satellite utilizes a 50 cm aperture telescope with a dichroic beam splitter that enables simultaneous observations in the FUV ($\lambda_{eff}$=1516\AA, FWHM=269\AA) and the NUV ($\lambda_{eff}$=2267\AA, FWHM=616\AA).  Martin et al. (2005), Morrissey et al. (2005), Morrissey et al. (2007) provide full details on the satellite, telescope, instrument and calibration and data processing pipeline.  Deep, single orbit ($\sim$1500 sec) imaging was obtained for each galaxy, following the strategy of the GALEX Nearby Galaxy Survey (Gil de Paz et al. 2007).  GALEX observations for a significant fraction of the remaining galaxies beyond these limits have been taken by other programs and are publicly available through MAST.\footnote{MAST is the Multimission Archive at the Space Telescope Science Institute (http://archive.stsci.edu/index.html).} Paper II provides photometry (following the methods of Gil de Paz et al. 2007) based upon both the archival data and 11HUGS imaging for over 85\% of the overall Paper I sample, and the currently available measurements are incorporated into our analysis.  Table 1 lists the $\sim$300 galaxies included in this study.

\section{Comparison of H$\alpha$ and FUV SFRs}

To begin, we compute the FUV and H$\alpha$ SFRs, where we correct for Milky Way foreground reddening but not for attenuation internal to the galaxies.  The standard conversion prescriptions of K98, where the ratio of the luminosities to the SFR are constants, are assumed.  We first describe the calculation of the SFRs, and then examine the consistency between the two tracers.

Corrections for foreground reddening are calculated by using $E(B-V)$ values based on the maps of Schlegel et al. (1998) and the Cardelli et al. (1989) extinction law with $R_V=3.1$.  The resulting relationships between the color excess and extinction 
 are $A_{FUV}=7.9E(B-V)$ and $A_{H\alpha}=2.5E(B-V)$.  

We compute H$\alpha$ SFRs using the emission-line fluxes given in Table 3 of Paper I.  The narrowband filters used for the survey are wide enough to capture the [NII]$\lambda\lambda$6548,83 lines which flank H$\alpha$, so the [NII] component of the observed flux needs to be removed before SFRs can be calculated.  Measurements of [NII]/H$\alpha$ from spectroscopy are available for about a quarter of the sample and are used to correct the fluxes in those cases.  A compilation of measurements from the literature is provided in Table 3 of Paper I.  Otherwise, the ratio is estimated from the average relationship between [NII]$\lambda$6583/H$\alpha$ and $M_B$, a consequence of the luminosity-metallicity relationship for galaxies.  The adopted scaling relationship, derived using the integrated spectroscopic dataset of Moustakas \& Kennicutt (2006), is:
\begin{equation}
\begin{array}{rl}
\mbox{log}(\mbox{[NII]}\lambda6583/\mbox{H}\alpha) = (-0.173\pm0.007)\;M_B - (3.903\pm-0.137) &\mbox{if}\; M_B > -20.3\\
\mbox{[NII]}\lambda6583/\mbox{H}\alpha =0.54 & \mbox{if}\; M_B \leq -20.3
\end{array}
\end{equation}

\noindent The total [NII]$\lambda\lambda$6548,83/H$\alpha$ ratio is computed assuming a 3-to-1 ratio between [NII]$\lambda$6583 and [NII]$\lambda$6548 (Osterbrock 1989).  Further details can be found in Appendix B of Paper I.  As noted there, although estimates based on this correlation are only good to a factor of $\sim$2, the majority ($>$70\%) of galaxies in our dwarf-dominated sample have $M_B>-17$, and hence have [NII]/H$\alpha$ ratios $\lesssim$0.1.  The corresponding uncertainties in the H$\alpha$ flux due to [NII] corrections are thus lower than $\lesssim$10\%. 

FUV SFRs are based on the FUV integrated photometry in Paper II, Table 2.  $f_{\nu}$ follows directly from the tabulated AB magnitudes given there since $f_{\nu}$[ergs s$^{-1}$ cm$^{-2}$ Hz$^{-1}$] $=10^{-0.4(m_{AB}+48.6)}$.  
Calculations based on the NUV data will generally be less reliable for tracing the recent star formation since the flux at these redder NUV wavelengths will have a greater contribution from stars with lifetimes $>$100 Myrs.

FUV and H$\alpha$ luminosities then follow from the distances in Paper I, Table 1, and the SFRs are computed as in K98:

\begin{equation}
  \mbox{SFR} [\mbox{M}_{\odot} \mbox{yr}^{-1}] = 7.9 \times 10^{-42}\; L(\mbox{H}\alpha) \; [\mbox{ergs s}^{-1}]
 \end{equation}
\begin{equation}
  \mbox{SFR} [\mbox{M}_{\odot} \mbox{yr}^{-1}] = 1.4 \times 10^{-28}\; L_{\nu}(\mbox{UV}) \; [\mbox{ergs s}^{-1} \mbox{Hz}^{-1}]
\end{equation}

\noindent The K98 conversion factors are calculated using a Salpeter IMF with mass limits of 0.1 and 100 M$_{\odot}$, and stellar population models with solar metal abundance.  The H$\alpha$ and UV calibrations also assume that the star formation history is constant for at least the past $\sim$10 Myrs and $\sim$100 Myrs, respectively.  Table 1 provides the computed SFRs along with some general properties of the galaxies in the sample.  The assumptions underlying the conversions and their validity in the context of the our dwarf galaxy dominated Local Volume sample are discussed further in the next section.  

\begin{figure}
\epsscale{1}
\plotone{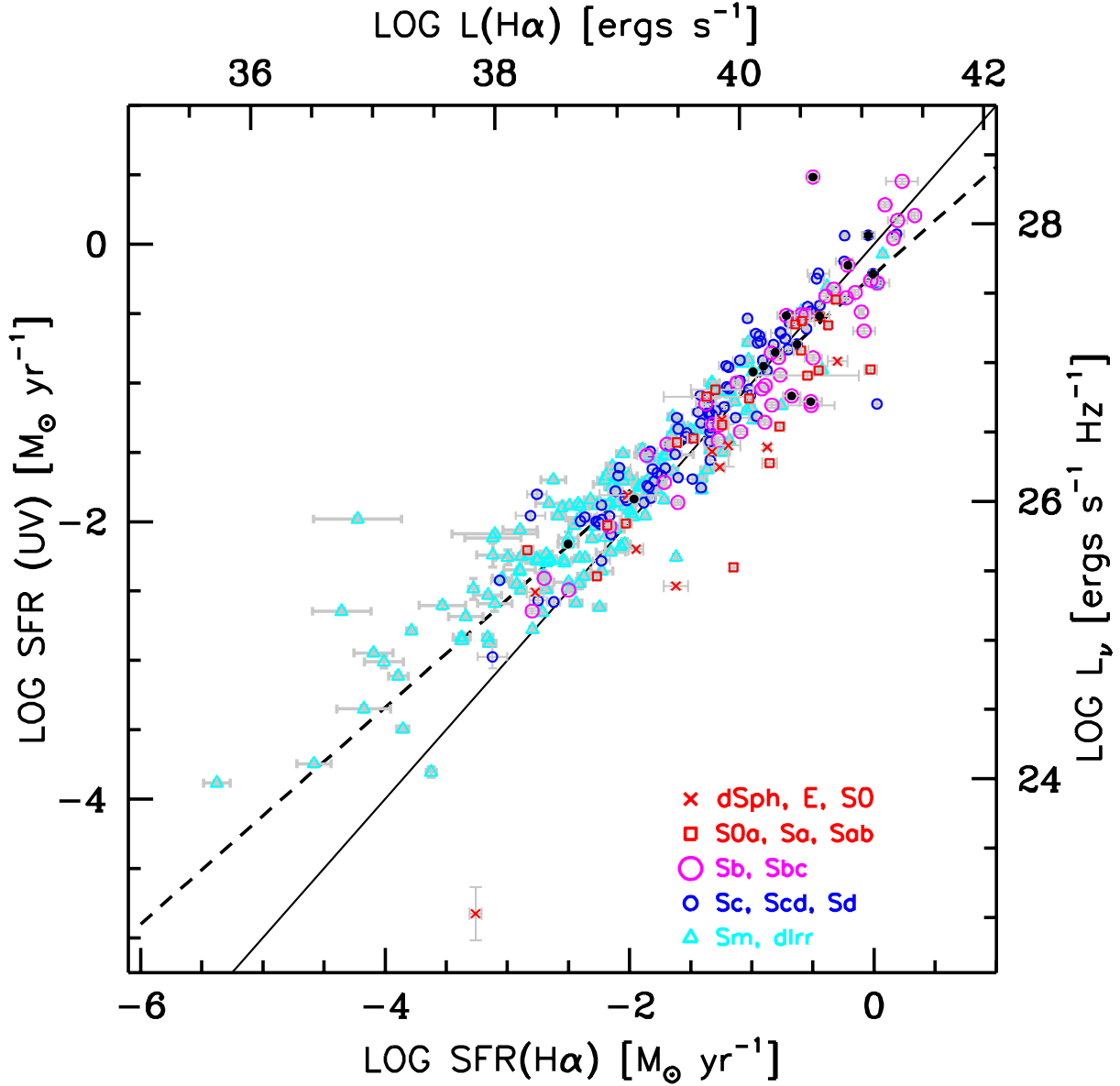}
\caption{Comparison of observed FUV and H$\alpha$ SFRs (i.e., without corrections for internal dust attenuation), calculated using the linear conversion recipes given in Kennicutt (1998).  Axes that indicate the corresponding H$\alpha$ and FUV luminosities are also shown.  Morphological type is distinguished with different symbols as indicated.  The solid line represents a one-to-one correspondence between the SFRs, while the dashed line is an ordinary least squares bisector fit to the data, given by log[SFR(FUV)]=0.79 log[SFR(H$\alpha$)] - 0.20. The black filled symbols represent galaxies where the H$\alpha$ flux may be underestimated because the narrowband imaging did not wholly enclose the galaxy and are excluded from the fit.}\label{fig:uvhacomp}
\end{figure}

In Figure~\ref{fig:uvhacomp}, we plot these non-dust-corrected FUV SFRs against the H$\alpha$ SFRs.  
The solid line represents a one-to-one correlation between the SFRs.  Axes indicating the corresponding luminosities are also shown.  Different colors and symbols are used to distinguish between morphological types as specified in the figure.  The error bars reflect uncertainties in the flux measurements, as described in Papers I and II.  A wide range of integrated SFRs are covered --- the sample probes normal spiral disks with SFRs of about unity to dwarf irregular galaxies with ultra-low SFRs of $\sim$10$^{-4}$ \msunyr.  

In general, there is coarse agreement between the FUV and H$\alpha$ SFRs.\footnote{There is one significant outlier at low luminosities where the H$\alpha$ SFR appears to exceed the FUV SFR by more than a factor of 10.  This system, KDG61, is in the M81 group, and Croxall et al. (2009) speculate that the system is a chance superposition of a dwarf spheroidal galaxy and an HII region from the M81 tidal stream.  Such a scenario can plausibly result in a high H$\alpha$-to-UV ratio and the extremely red color (FUV-NUV$>$2, which is not the result of extinction since H$\alpha$/H$\beta$ is close to the expected Case B recombination value) that are observed.  This data point is excluded from the least squares fits that follow.}  
The SFRs agree to within about a factor of 2 for the majority of galaxies with SFR$\gtrsim$0.01 \msunyr, even before internal extinction corrections are applied.  This agreement is reasonable given the relative absence of highly-obscured, luminous galaxies in the small volume probed by our sample.  However, there is a clear systematic present: 
at low SFRs (log SFR$\sim$-1.5; SFR$\lesssim$0.03\msunyr) H$\alpha$ begins to underpredict the SFR relative to the FUV luminosity.  A linear least squares bisector fit to the data demonstrates this tilt (dashed line)\footnote{The filled black symbols represent galaxies where the H$\alpha$ flux may be underestimated because the narrowband imaging did not wholly enclose the galaxy.  These points are excluded from the fit, and omitted from Table 1.  In the next figure, these same galaxies are plotted with the circled points, and are also excluded from the fit and reported statistics.}.  

\begin{figure}
\epsscale{1}
\plotone{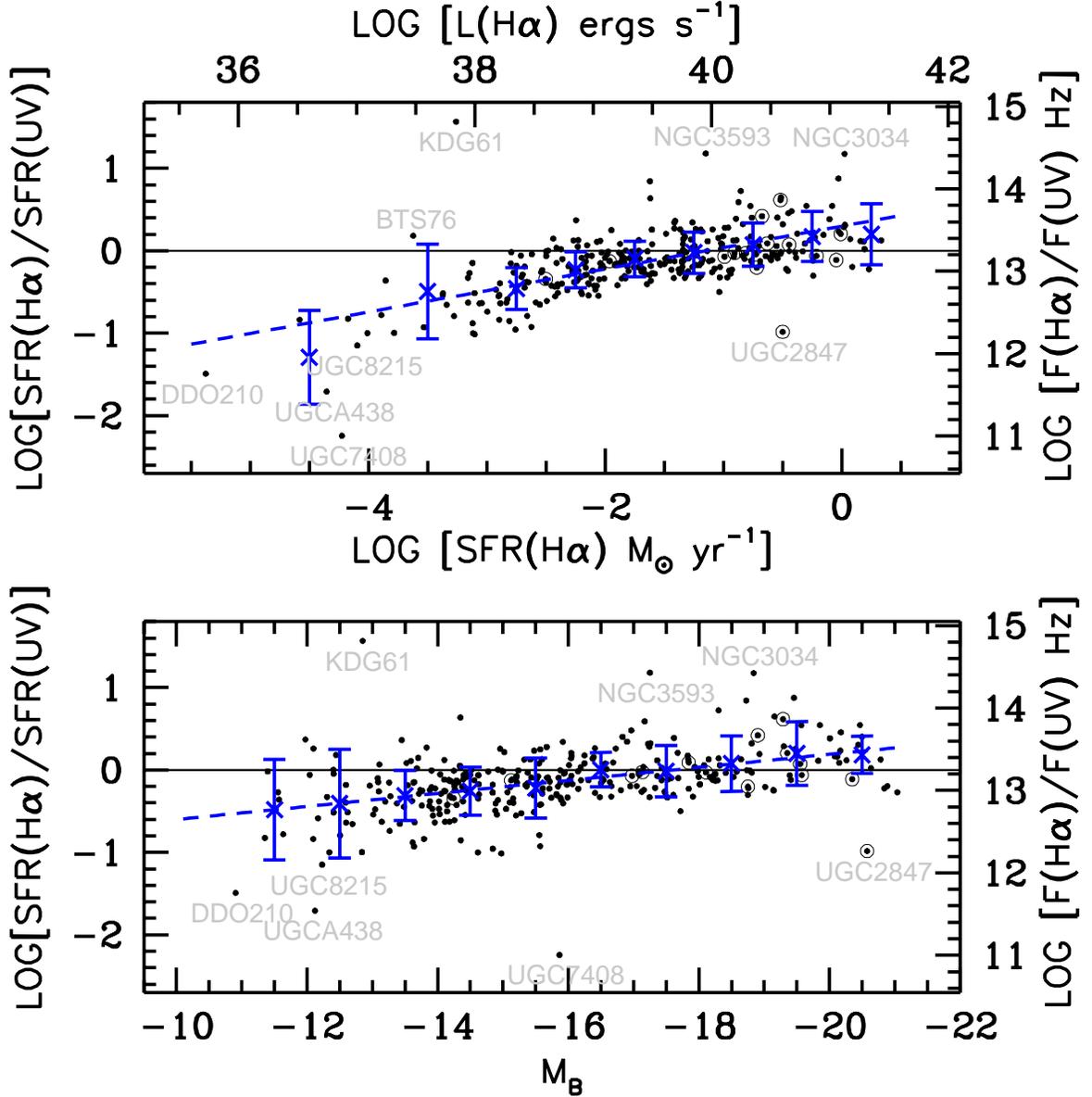}
\caption{The ratio of H$\alpha$-to-FUV SFRs, where corrections for internal dust attenuation have not yet been applied (or equivalently, the observed H$\alpha$-to-FUV flux ratio), as a function of the H$\alpha$ SFR/luminosity (top panel) and the B-band absolute magnitude (bottom panel).  The solid lines indicate ratios of unity, while the dashed line shows the linear least squares fits to the data, which are given by log[SFR(H$\alpha$)/SFR(FUV)]=0.26 log[SFR(H$\alpha$)]+0.30, and log[SFR(H$\alpha$)/SFR(FUV)]=-0.08 $M_B$ -1.39.  Binned averages of the data with 1$\sigma$ scatter are also overplotted (blue symbols) and are listed in Table 2. Circled points represent galaxies where the H$\alpha$ flux may be underestimated because the narrowband imaging did not wholly enclose the galaxy, and are excluded from the fit and statistics.}\label{fig:uvharesidual}
\end{figure}

The systematic offset is better illustrated in the top panel of Figure~\ref{fig:uvharesidual}  which plots the ratio of the SFRs as a function of the H$\alpha$ SFR.  The covariance between the axes acts to emphasize the deviation from unity.  The right and top axes indicate the corresponding flux ratios and luminosities, respectively.  Here, a linear least squares fit to the data is shown (dashed line) and has a scatter of 0.30 dex.  Binned averages with 1$\sigma$ ranges are also overplotted (blue symbols), and the values are given in Table~\ref{tab:binnedavgs}.  For L(H$\alpha$)$\gtrsim10^{40}$ ergs s$^{-1}$ (SFR$\gtrsim$ 0.1\msunyr) there is good consistency between the two tracers, and the average deviation from unity is 30\%.  For L(H$\alpha)\lesssim 4 \times 10^{38}$ ergs s$^{-1}$ (SFR$\lesssim$ 0.003\msunyr), H$\alpha$ yields SFRs that are lower than the FUV by average factors ranging from two to more than ten.  An analogous plot as a function of the B-band absolute magnitude is shown in the bottom panel of Figure~\ref{fig:uvharesidual}.  Here, the tilt in the correlation is less steep, and the scatter is larger (0.35 dex).  For galaxies brighter than $-$19 mag the H$\alpha$ yields an SFR that is $\sim$50\% higher.  There is good consistency for $-16\gtrsim M_B \gtrsim-19$, and average offsets of a factor of 2 and higher begin to appear at $\gtrsim-$14 mag.

\section{Discussion: Understanding the Systematic Decline in L(H$\alpha$)/L(FUV)}

Using a complete, statistical sample of star-forming galaxies within the Local Volume, we find that the observed H$\alpha$-to-FUV flux ratio systematically decreases with declining SFR and galaxy luminosity.  When standard conversion recipes of the form SFR/L=constant are applied (e.g., K98), the H$\alpha$ luminosity will thus underestimate the SFR relative to the FUV luminosity in dwarf galaxies.  This analysis confirms earlier indications of a such a trend for L(H$\alpha$)$\lesssim10^{40}$ or SFR$\lesssim$0.1 \msunyr (see introduction for references).  Our more robust sampling of low-mass star-forming galaxies represents a significant improvement over past studies which were only able to probe to regimes of $-2\lesssim$ log SFR(H$\alpha$)$\lesssim-1$ with $\sim$20 dwarf galaxies (c.f. Sullivan et al. 2000, Figure 14; Bell \& Kennicutt 2001, Figure 2).  The recent study of Meurer et al (2009), which has examined correlations in the H$\alpha$-to-FUV flux ratio with H$\alpha$ and $R$-band surface brightness, also only includes $\sim$25 galaxies with $-3\lesssim$ log SFR(H$\alpha$)$\lesssim-1$ (M. Seibert, private communication).  In contrast, the current analysis is based on over 200 dwarf galaxies with SFR$\lesssim$0.1 \msunyr\ and extends to ultra-low activities of SFR$\sim 10^{-4}$ \msunyr.  We find that H$\alpha$ SFRs of 3 $\times 10^{-3}$ \msunyr\ are lower than those inferred from the FUV by a factor of $\sim$2 and the discrepancy increases to a factor of $\gtrsim$10 by SFR(H$\alpha$)=10$^{-4}$ \msunyr, on average.

The expectation that the two tracers should yield consistent SFRs relies on the robustness of the SFR calibrations that are used.  In the K98 formulation, consistent SFRs require that L(H$\alpha$)/L(FUV)=constant, since SFR/L(H$\alpha$) and SFR/L(FUV) are also both simple constants.  Therefore, a convenient framework for understanding the observed deviations from the expected ratio involves checking the validity of the assumptions underlying the calibrations with respect to the particular galaxies to which they are being applied.  To briefly summarize how the constants are derived, stellar population synthesis models are used to compute H$\alpha$ and UV luminosities, given an invariant stellar initial mass function (IMF).  The expected flux is calculated by integrating a constant SFR over a time period that is long enough to allow the birth and death of the stars responsible for the emission to equilibrate.  Solar metallicity models are normally used, and the IMF is well-populated at all masses.\footnote{For example, the Starburst99 models of Leitherer et al. (1999) use 10$^6$ M$_{\odot}$ single-age stellar populations to construct systems with more complex star formation histories, and such building blocks contain $\sim$4500 O-stars at zero-age for a Salpeter IMF with mass limits of 1 and 100 \msunyr.  For more commonly used mass limits of 0.1 and 100 \msunyr, the corresponding number of O-stars would be lower, but still well sampled, at $\sim$2500.}  The nebular emission luminosity is computed assuming Case B recombination to convert the production rate of ionizing photons to an H$\alpha$ luminosity.  Finally, the model-predicted luminosities are typically intrinsic values that have not yet been attenuated by dust.

All of these ingredients have been examined in previous studies to investigate possible trends in the L(H$\alpha$)/L(FUV) ratio, but primarily in the context of more massive star-forming systems than typical of our Local Volume sample.  The effects of uncertainties in the stellar evolution tracks and model atmospheres, non-solar metallicities, non-constant SFHs, leakage of ionizing photons, departures from Case B recombination, dust attenuation, stochasticity in the formation of high-mass stars, and finally, variations in the IMF have all been considered (e.g., Buat et al. 1987; Buat 1992; Sullivan et al. 2000; Bell \& Kennicutt 2001; Buat et al. 2002; Iglesias-Paramo et al. 2004; Sullivan et al. 2004; Salim et al. 2007) and have been recently revisited by Meurer et al. (2009).  More mundane measurement systematics such as photometric aperture mismatches between the FUV and H$\alpha$ data have also been considered\footnote{The H$\alpha$ fluxes given in Paper I and the FUV magnitudes given in Paper II that are used in the SFR comparison analysis represent integrated measurements which are not necessarily taken through the same apertures.  However, we have checked that the impact of this on our analysis and conclusions is negligible.  This is because the GALEX and H$\alpha$ apertures cover the same general area for most galaxies.  There are some cases, however, where the H$\alpha$ aperture is smaller because (i) it is limited by the FOV of the optical detector or (ii) the FUV emission is more extended.  In such cases, GALEX fluxes were re-measured using the apertures used for the H-alpha photometry and there are no significant differences in the correlations when these fluxes are used.} (e.g., Sullivan et al. 2000).  We re-evaluate these issues in the context of our dwarf-dominated sample, and tackle the ones which are more straightforward to examine first.

\subsection{Internal Dust Attenuation}
\label{sec:dustcorr}

In general, attenuation by dust is the first issue to address when comparing 
UV/optical observed luminosities with intrinsic model values.  For normal spiral galaxies (SFR$\gtrsim$1 \msunyr) there is broad agreement among previous work that although the UV SFR tends to be lower than the H$\alpha$ SFR before accounting for internal dust attenuation, there is good consistency between the two after proper corrections have been applied.  Empirically, these results suggest that the UV stellar continuum is more affected by attenuation than the H$\alpha$ nebular emission\footnote{Based on extinction curves alone, it may seem obvious that the UV should be more attenuated than redder emission at 6563\AA.  However, differential reddening between gas and stars must also be considered.  As discussed in Calzetti et al. (1994), the nebular emission (from HII regions which are the sites of the most recent star formation and are enshrouded by dust) suffers about twice the reddening experienced by the stellar continuum, and this leads to an H$\alpha$ attenuation that is more comparable to that of the UV continuum.}.  If the same holds true for lower mass dwarf galaxies, dust corrections would drive the H$\alpha$-to-UV flux ratio even lower, which would at first appearance seem to exacerbate rather than help reconcile the discrepancy.  To check this for our sample, we estimate the attenuation in the H$\alpha$ flux with the Balmer decrement, and in the FUV flux using the total infrared-to-UV (TIR/FUV) flux ratio.  The two methods are independent, the former being rooted in recombination physics and the latter in stellar synthesis and dust models, although an extinction curve must be assumed in each case.  The resulting dust corrections are provided in Table 1.

\subsubsection{$A_{H\alpha}$ via the Balmer Decrement}
The validity of using the Balmer decrement to infer the average amount of nebular extinction in galaxies is supported by studies that have shown that the scatter between infrared-based SFRs and H$\alpha$ SFRs is decreased when the H$\alpha$ luminosity is corrected for extinction in this manner (e.g., Dopita et al. 2002; Kewley et al. 2002; Moustakas et al. 2006).  Kennicutt et al. (2009) reaches similar conclusions from the analysis of about 400 nearby galaxies.  As in Lee (2006), we assume an intrinsic Case B recombination ratio of 2.86 for H$\alpha$/H$\beta$, and the Cardelli et al. (1989) Milky Way extinction curve with $R_V=3.1$, to express $A_{H\alpha}$ in terms of the observed H$\alpha$/H$\beta$ ratio as:

\begin{equation}
A_{H\alpha}= 5.91\;\mbox{log}\frac{f_{H\alpha}}{f_{H\beta}} - 2.70.
\end{equation}

\begin{figure}
\epsscale{1}
\plotone{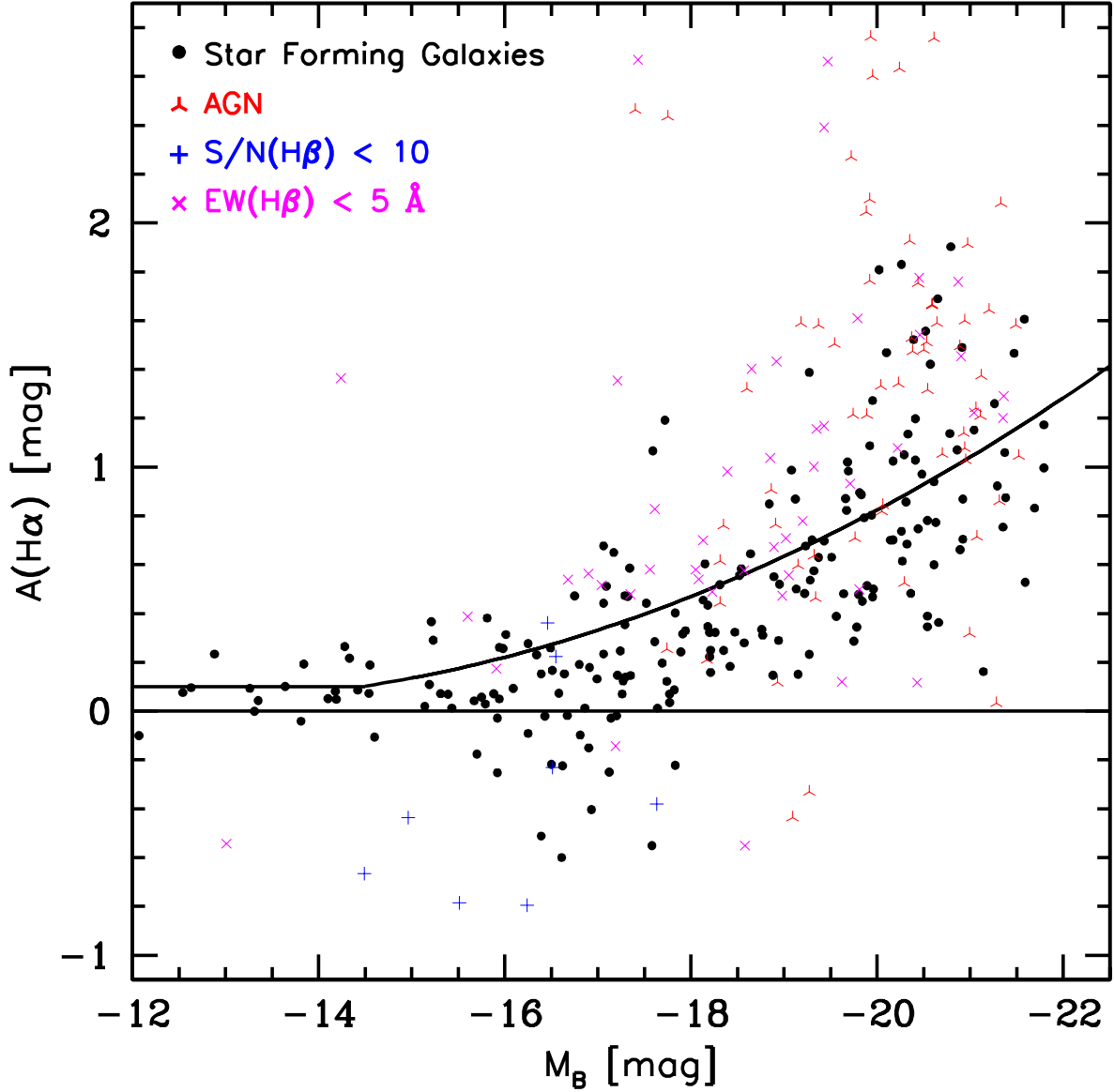}
\caption{$A({H\alpha})$, the H$\alpha$ nebular dust attenuation in magnitudes, based on measurements of the Balmer decrement, plotted against the B-band absolute magnitude $M_B$ for galaxies in the integrated spectral atlas of Moustakas \& Kennicutt (2006).  The black filled circles are star-forming galaxies, and those with $A({H\alpha})>0$ have been included in the over-plotted fit (see Equation~\ref{eq:Aha_MB}).  The various crosses represent galaxies which have been excluded from the fit because they have line ratios indicating the presence of an AGN (red), signal-to-noise in H$\beta<$10 (blue), or H$\beta$ equivalent width $<$ 5\AA\ (magenta).  Luminous star-forming galaxies exhibit greater levels of nebular attenuation than those at lower luminosity.  The over-plotted fit is used to estimate the nebular extinction when measurements of the Balmer decrement from integrated spectral measurements are not available.}\label{fig:haattenuation}
\end{figure}

For $\sim$20\% of the sample, spectroscopic measurements of H$\alpha$/H$\beta$ from the literature are available. 
For those without robust spectroscopic measurements, we follow a strategy similar to that used for estimating [NII]/H$\alpha$ above.  Using the integrated spectroscopy of Moustakas \& Kennicutt (2006) we derive an empirical scaling relation between $A_{H\alpha}$ and $M_B$ which is shown in Figure~\ref{fig:haattenuation}.  
The data exhibit the well-known trend that the more luminous galaxies are more heavily obscured.  We fit a piecewise function to the data, described by a constant value at the lowest luminosities and a second order polynomial for $M_B<-15.0$.  Objects that have $A_{H\alpha}<0$, a signal-to-noise less than 10 in the H$\beta$ line, an H$\beta$ equivalent width $<5$\AA, or show evidence of AGN emission are excluded from the fit.  The adopted scaling relation is:

\begin{equation}\label{eq:Aha_MB}
A_{H\alpha}=\left\{ \begin{array}{lll}
0.10 \;&\mbox{if}\; M_B > -14.5\\
1.971+0.323M_B+0.0134M_B^2 \;& \mbox{if}\; M_B \leq-14.5\\
\end{array}
\right.
\end{equation}

Of course, there is a great deal of scatter from this average relationship which becomes more severe with increasing luminosity.  However, although there is a 40\% scatter for galaxies with $M_B<-18$, the scatter is much lower for the relatively transparent dwarf galaxies which dominate our Local Volume sample: for $-14.7>M_B>-18$ it is 20\% and for the lowest luminosity galaxies, where we have assumed a constant average correction, it drops further to 10\%.  

\subsubsection{$A_{FUV}$}
The total infrared-to-UV flux ratio (TIR/FUV) is an indicator of the UV attenuation in a galaxy as it represents the total stellar light which has been absorbed and re-radiated by dust, relative to the surviving stellar UV light which is directly observed.  Using stellar population models with various dust geometries and extinction curves, it has been shown that for systems with recent star-formation, such as the spiral and irregular galaxies in our sample, the mapping between A(FUV) and TIR/FUV is relatively tight ($\sim 20\%$); that is, the contribution of less massive/older stellar populations to the heating of the dust is either negligible or is in direct proportion to that of the O\&B stars which dominate the UV emission (e.g., Buat et al. 2005).  The mapping as given by Buat et al. is:

\begin{equation}\label{eq:buat}
A(FUV) = -0.0333 x^3 + 0.3522 x^2 + 1.1960 x^3 +0.4967
\end{equation}

\noindent where x=log($\frac{TIR}{FUV}$).
MIPS 24, 70 and 160$\mu$m integrated fluxes from the Spitzer LVL program are available for the majority of galaxies used in this analysis (Dale et al. 2009), and we compute TIR fluxes for those galaxies that have robust detections in all three bands (50\% of our sample) using the calibration of Dale \& Helou (2002):

\begin{equation}\label{eq:dale}
TIR=1.559 \nu f_{\nu}(24)+ 0.7686 \nu f_{\nu}(70)+ 1.347 \nu f_{\nu}(24).
\end{equation}

\noindent FUV is computed as $\nu f_{\nu}$ at 1520\AA.  

\begin{figure}
\epsscale{1}
\plotone{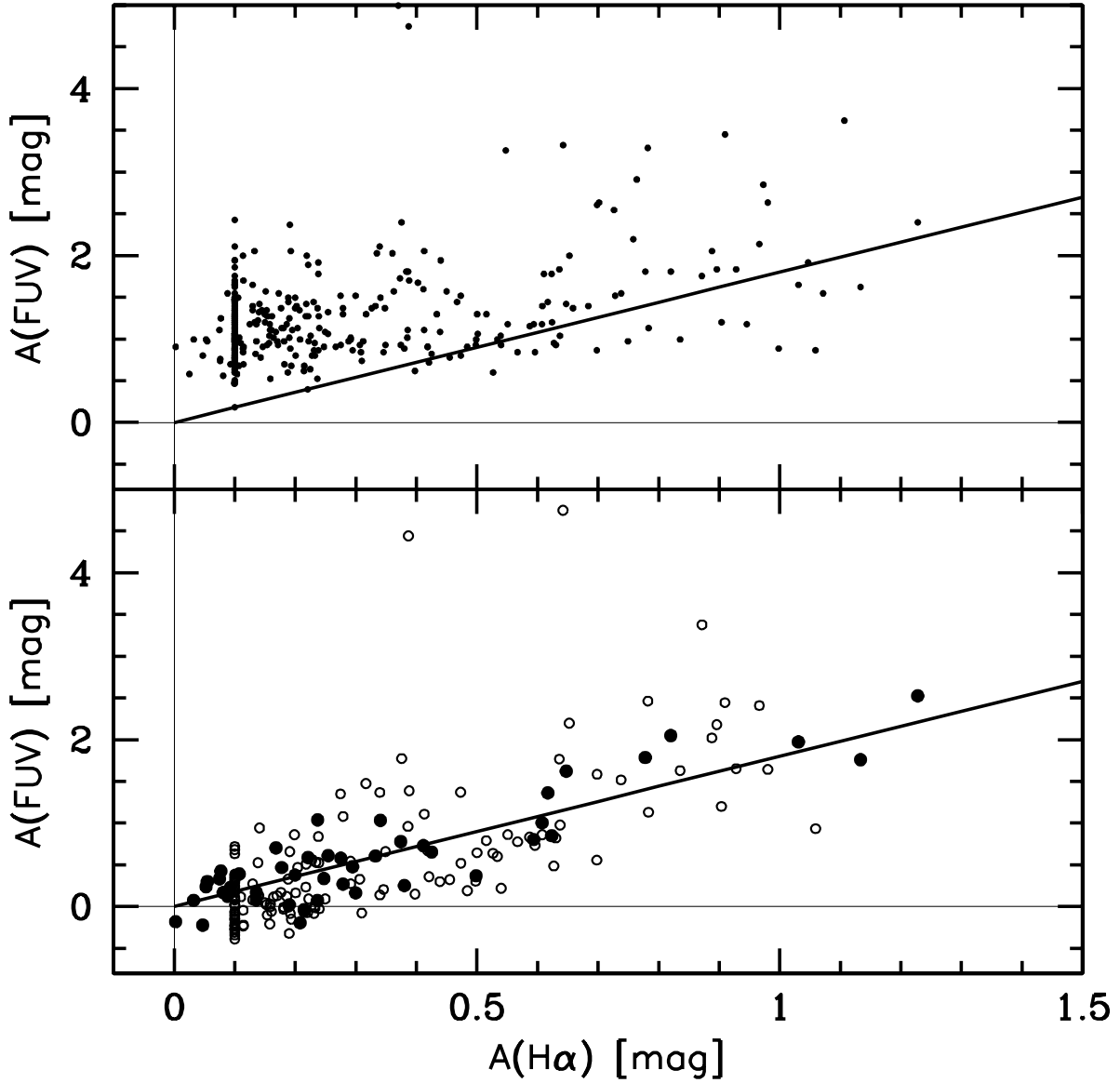}
\caption{A comparison of different estimates of the H$\alpha$ and FUV attenuation in galaxies in the 11HUGS sample.  {\it Bottom panel:} FUV attenuations, A(FUV), based on TIR/FUV ratios computed from MIPS photometry, and H$\alpha$ attenuations A(H$\alpha$) based on (i) integrated spectroscopic measurements of the Balmer decrement (solid large points) and, (ii) A(H$\alpha$) estimated from a $M_B$ dependent scaling relationship calibrated with Balmer decrement measurements (open points). {\it Top panel:} A(FUV) based on the linear IRX-$\beta$ relationship of Cortese et al. (2006) and our best available value for A(H$\alpha$) (small black points).  These latter A(FUV) estimates appear to be too high and are not used in our analysis.  The solid line in both panels represents A(FUV)=1.8A(H$\alpha$), as expected for the Calzetti obscuration curve and differential extinction law.  The Calzetti law provides a good description of our most robust attenuation measurements (solid large points). }\label{fig:attenuationcomp}
\end{figure}

For those galaxies for which both a spectroscopic measurement of H$\alpha$/H$\beta$ and MIPS FIR data are available, we compare the H$\alpha$ and FUV attenuations in bottom panel of Figure~\ref{fig:attenuationcomp} (solid points).  There is a good correlation with a slope of 1.8 (solid line).  Interestingly, A(FUV)/A(H$\alpha$)=1.8 is what is expected for the Calzetti obscuration curve and differential extinction law (Calzetti 2001), which is based on UV luminous starbursts.  This agreement provides some assurance that the attenuation corrections we have derived are reasonable and generally consistent.  In Figure~\ref{fig:attenuationcomp} we also plot the A(H$\alpha$) estimated from $M_B$ using equation \ref{eq:Aha_MB} (open circles).  As expected, there is more scatter between A(FUV) and the estimated A(H$\alpha$), but the values are still correlated with a slope of $\sim$1.8.

When FIR data are not available, it is common to resort to methods which estimate A(FUV) directly from some measure of the UV spectral slope (e.g., a UV color).  Previously, this method appeared to be reasonable since it is well-known that the UV spectral slope $\beta$ and TIR/FUV are tightly correlated (and thus the UV slope and  A(FUV) should also be well-correlated), at least for samples of local starbursts (Calzetti et al. 1994; Meurer et al. 1999).  It has become clear however that the ``IRX-$\beta$'' correlation for starbursts does not hold for normal star-forming galaxies, which have lower values of TIR/FIR at a given UV color, and more scatter between the quantities.  Calibrations and scaling relationships which better describe the correlation for such systems have been published (e.g., Kong et al. 2004; Seibert et al. 2005; Cortese et al. 2006; Boisser et al. 2007; Gil de Paz et al. 2007).  However, as we show in Dale et al. (2009) and Lee et al. (2009b), the majority of galaxies in our local volume sample are low-luminosity dwarf irregulars which are blue (0$\lesssim$FUV-NUV$\lesssim$0.5), have low TIR/FUV ($\lesssim2$), and thus lie in the area of the IRX-$\beta$ diagram where the UV color is not (strongly) correlated with TIR/FUV and attenuation.  Thus, such IRX-$\beta$ relations cannot be used to infer TIR/FUV and A(FUV) for our sample overall.  For example, a relation by Cortese et al. (2006), which is based on normal but more luminous spirals and provides a poor fit for galaxies with TIR/FUV$\lesssim2$ results in A(FUV) estimates that are too high when compared to the attenuations based on the Balmer decrement, as illustrated in Figure~\ref{fig:attenuationcomp}.  We thus take another approach for estimating A(FUV) for galaxies with no FIR data or only upper-limit detections.  Based on Figure~\ref{fig:attenuationcomp}, we simply scale the computed A(H$\alpha$) values (whether calculated from the Balmer decrement or estimated from $M_B$) by a factor of 1.8.  This value is also adopted when Equations~\ref{eq:buat} and~\ref{eq:dale} produce a negative correction (i.e., for TIR/FUV$<$0.3).  In all, A(FUV) is estimated from A(H$\alpha$) for $\sim$60\% of the galaxies in Table 1.  Based on comparison with TIR/FUV based attenuations (Figure~\ref{fig:attenuationcomp}), there is a 1$\sigma$ uncertainty of $\sim$17\% when estimating A(FUV) from Balmer decrement based A(H$\alpha$) values.  When A(H$\alpha$) is itself estimated (equation 5), the average uncertainties increase (26\%), and are larger for more luminous objects ($M_B<-18;\sim$40\%) than for the dwarf galaxies that dominate our local volume sample ($M_B>-18;\sim$20\%).

\subsubsection{Comparison of Dust Corrected SFRs}

\begin{figure}
\epsscale{1}
\plotone{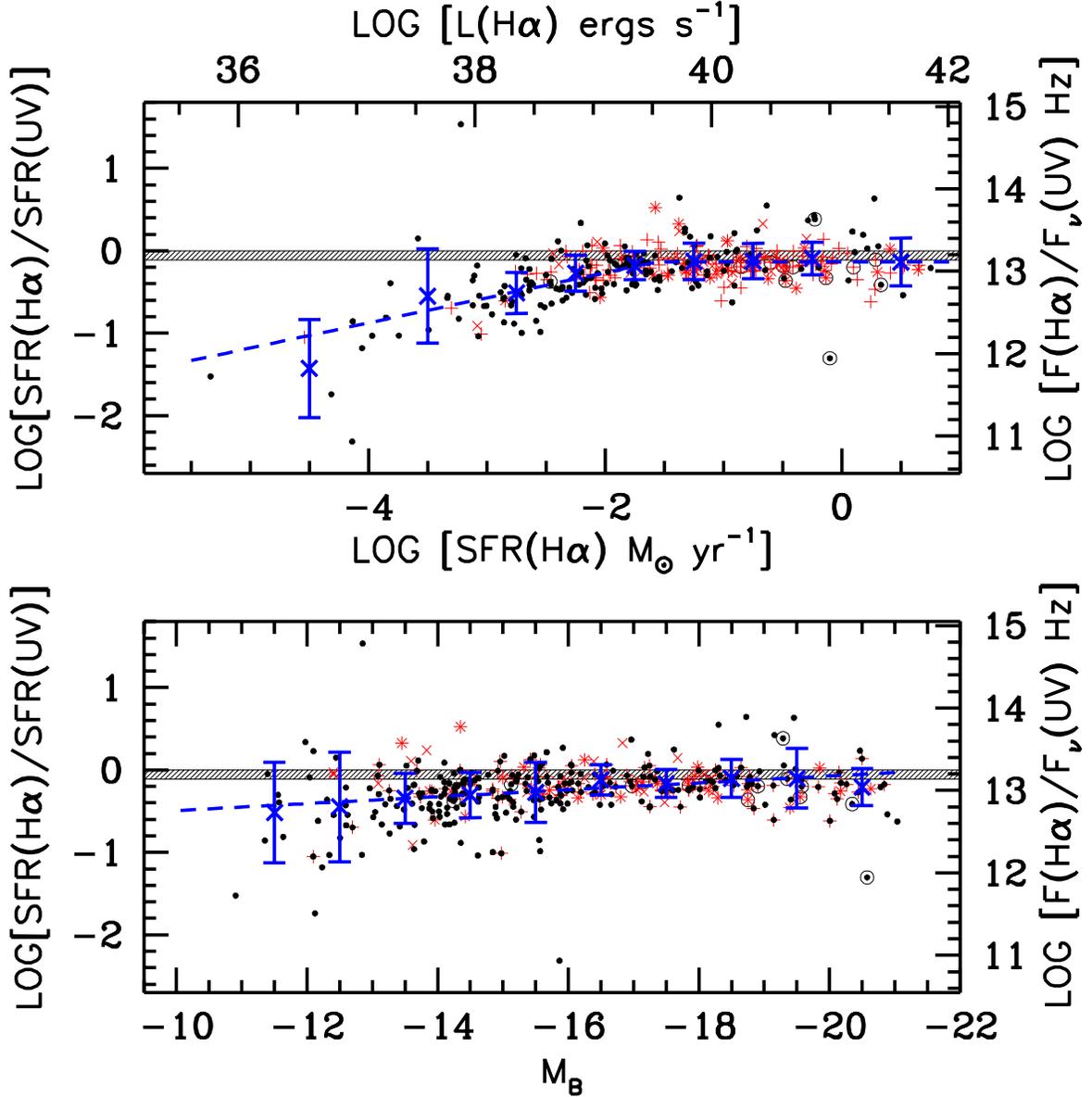}
\caption{Same as Figure \ref{fig:uvharesidual}, but corrected for attenuation by dust internal to the galaxies as described in Section~\ref{sec:dustcorr}. Data points which have nebular attenuations computed from the Balmer decrement are distinguished by red X's, and those where the FUV attenuations are based on the TIR-to-FUV ratio are shown as red crosses.  Galaxies for which both Balmer decrement and TIR-to-FUV based corrections are applied thus appear as red stars.  Corrections for the remaining points are estimated with scaling relationships.  The shaded band represents the range of H$\alpha$-to-FUV ratios predicted by commonly used stellar population models as described in Section~\ref{sec:models}.  The dashed lines in the top panel are given by log[SFR(H$\alpha$)/SFR(FUV)]=-0.13 for log[SFR(H$\alpha$)]$>$-1.5, and log[SFR(H$\alpha$)/SFR(FUV)]=0.32 log[SFR(H$\alpha$)]+0.37 for log[SFR(H$\alpha$)]$<$-1.5.  In the bottom panel the fit is given by log[SFR(H$\alpha$)/SFR(FUV)]=--0.05 $M_B$ --0.99.}\label{fig:uvharesidualdustcor}
\end{figure}

After accounting for the effects of dust using the best available attenuation value for each galaxy, the H$\alpha$-to-FUV flux and SFR ratios are replotted in Figure~\ref{fig:uvharesidualdustcor}.  Data points where more robust corrections are applied (i.e., based on Balmer decrements and/or TIR-to-FUV ratios) are distinguished in red.  There are no significant differences between the trends described by galaxies where scaling relationships are used to estimate attenuations and those which have more robust corrections.  Binned averages with 1$\sigma$ scatters are again given in Table~\ref{tab:binnedavgs}.  For all galaxies, the correction increases the FUV flux relative to the H$\alpha$ flux (FUV is more attenuated as expected).  However, since higher luminosity galaxies tend to suffer from more attenuation than those at lower luminosity (e.g., Figure~\ref{fig:haattenuation}), the H$\alpha$-to-FUV flux ratio is depressed by a greater factor at the high luminosity end.  At L(H$\alpha$)$\sim10^{41}$ ergs s$^{-1}$ (SFR $\sim$ 1\msunyr), the ratio decreases by $\sim$ 0.2 dex, and as a result falls below the K98 value (solid line).  Galaxies with L(H$\alpha$)$\lesssim10^{38}$ ergs s$^{-1}$ (SFR(H$\alpha) \lesssim$ 10$^{-3}$ \msunyr) are minimally affected.  The main consequence is that the slope above SFR(H$\alpha) \sim$ 3 $\times 10^{-2}$ \msunyr\ is flattened (i.e., the dust-corrected ratio in this regime is constant on average).  Adopting this lower value as the fiducial expected ratio (instead of the K98 value) would have the effect of mitigating the relative discrepancy at lower luminosities by 0.1 dex. 
Even in this situation however, factor of 2 offsets in the H$\alpha$-to-FUV ratio at SFR(H$\alpha) \sim$ 2 $\times 10^{-3}$ \msunyr, which increase to factors of $\gtrsim10$ at SFR(H$\alpha$)=10$^{-4}$ \msunyr, would still remain. 

\subsection{Stellar Model Uncertainties}\label{sec:models}

Differences in stellar evolution and atmosphere models used to calibrate H$\alpha$ and FUV luminosities as SFR indicators give rise to differences in the respective SFR conversion factors and hence to the expected H$\alpha$-to-FUV ratio.  While this will not produce systematic trends in the ratio as a function of the luminosity for a given metallicity and IMF, it does define the fiducial from which deviations are measured.  In Figure~\ref{fig:uvharesidualdustcor}, a gray shaded area is overplotted to indicate the range of ratios based on widely used synthesis models for solar metallicity and a Salpeter IMF with mass limits of 0.1 and 100 \msunyr.  These have been computed by Iglesias-Paramo et al. (2004) and Meurer et al. (2009) for the synthesis codes of Leitherer et al. (1999; Starbust99), Bruzual \& Charlot (2003; BC03) and Fioc \& Rocca-Volmerange (1997; PEGASE).  All of the models primarily adopt stellar evolutionary tracks from the Padova group (e.g., Girardi et al. 1996 and references therein), but differ in their treatment of stellar atmospheres.  There are uncertainties of $\sim$20\% in the H$\alpha$-to-FUV ratio due to the models alone.  The dust-corrected ratios computed in the last section for the more luminous galaxies in our sample are within the range of expected model values.  It thus appears reasonable to use galaxies with SFR(H$\alpha) \gtrsim$ 3 $\times 10^{-2}$ \msunyr\ to {\it empirically} define the fiducial H$\alpha$-to-FUV ratio at solar metallicity.  Hereafter, we measure deviations from there instead of the K98 value.

\subsection{Metallicity}
Standard SFR conversion recipes (and hence the expected H$\alpha$-to-FUV ratio) generally assume solar metallicity populations.  However, our sample spans a range of metallicities and is increasingly dominated by metal-poor dwarfs at low luminosities and SFRs.  Given this trend in sample properties, it is possible that the variation in the H$\alpha$-to-FUV ratio could potentially be driven by systematic variations in metallicity.  

Metallicity influences the SED (spectral energy distribution) through its effect on the stellar opacity. Lower metallicity stars of a given mass will have lower opacities, lower pressures, and thus will be relatively smaller and have hotter atmospheres.  They produce a larger number of UV photons (both ionizing and non-ionizing), so SFR conversions based on solar metallicity populations will tend to over-estimate the true SFR when applied to metal-poor systems (e.g., Lee et al. 2002, Brinchmann et al. 2004, Lee et al. 2009a).  In the same vein however, metal-poor populations will produce more ionizing flux relative to non-ionizing UV continuum, leading to {\it larger} H$\alpha$-to-FUV ratios at low metallicity.  This produces an effect which is the opposite of that observed, as also previously noted by Sullivan et al. (2000), Bell \& Kennicutt (2001) and Meurer et al. (2009), and therefore cannot be the cause of the discrepancy.

\begin{figure}
\epsscale{0.8}
\plotone{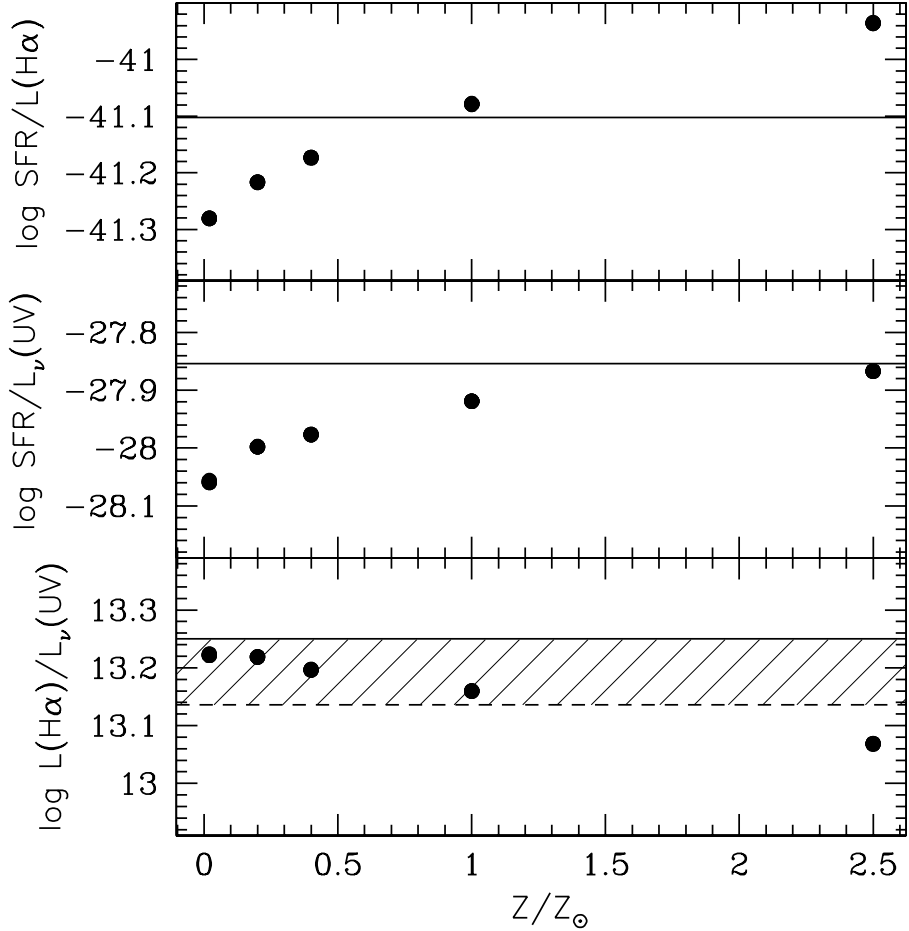}
\caption{Star formation rate conversion factors and the expected H$\alpha$-to-FUV ratio as a function of metallicity calculated from the Bruzual \& Charlot (2003) population synthesis models assuming a Salpeter IMF.  The top panel shows the H$\alpha$ conversion factor in units of log (M$_{\odot}$ yr$^{-1}$)/(ergs s$^{-1}$), assuming case B recombination with nebular temperatures and densities of 10$^{4}$ K and 100 cm$^{-3}$.  The middle panel shows the conversion factor for the FUV flux density in units of log(M$_{\odot}$ yr$^{-1}$)/(ergs s$^{-1}$ Hz), calculated using the GALEX FUV filter profile.  The bottom panel shows the corresponding expected ratios.  The solid horizontal line in each panel indicates the values as given in K98.  In the bottom panel, the shaded region shows the range of ratios based on widely-used synthesis models as described in Section~\ref{sec:models} and also plotted in Figure~\ref{fig:attenuationcomp}.  The H$\alpha$-to-FUV flux ratio increases with decreasing metallicity.}\label{fig:metallicity_models}
\end{figure}

We illustrate these effects in Figure~\ref{fig:metallicity_models}, which presents the SFR conversion factors and expected H$\alpha$-to-FUV ratio, based on the Bruzual \& Charlot (2003) stellar population synthesis models with Salpeter IMF and ``Padova 1994'' tracks.  Values ranging from Z$_{\odot}$/50 (characteristic of the nebular oxygen abundance of the most metal poor local galaxies known; Kunth \& Ostlin 2000) to 2.5Z$_{\odot}$ are shown.  Metallicity has less of an effect on the FUV than the H$\alpha$ (note that the vertical scales on all panels span the same range), resulting in larger H$\alpha$-to-FUV ratios at low metallicity.  Based on these models, the difference in the expected ratios at sub-solar metallicities relative to solar is small, $<$0.07 dex, so this should minimally impact the observed trends in Figure~\ref{fig:uvharesidualdustcor}.  We check on the magnitude of the effect by first estimating metallicities from the luminosity-metallicity relationship found locally for dwarf irregular galaxies (Lee et al. 2003), and then computing corrections for the SFRs of the individual galaxies in the sample.  Replotting the metallicity-corrected SFRs confirms that the effect on the magnitude of the overall systematic decrease in the H$\alpha$-to-FUV ratio is negligible.  It should be kept in mind, however, that the evolution of massive stars at sub-solar metallicity is currently ill-constrained due to uncertainties in stellar rotation and mixing which may amplify this effect (e.g., Gallart et al. 2005, Leitherer 2008 and references therein).

\subsection{Ionizing Photon Loss}
In the calibration used to convert the H$\alpha$ luminosity to the SFR, it is assumed that every Lyman continuum photon emitted results in the ionization of a hydrogen atom.  However, if there is leakage of Lyman continuum photons into the IGM, the observed flux would be lower than expected, and the true SFR would be underestimated from H$\alpha$. In order for leakage alone to explain the observed trend in the H$\alpha$-to-FUV ratio, the fraction of ionizing photons escaping the galaxies would have to increase with decreasing luminosity and SFR, and at least half of the ionizing photons would have to be lost in the typical system with log L(H$\alpha$)$\sim$38.4.  While the gas column densities in dwarf galaxies tend to be lower (e.g., Bigel et al. 2008), and there could be more leakage from the individual HII regions (the fraction of diffuse ionized gas appears to be nominally higher in dwarfs; e.g., Oey et al. 2007), star forming dwarf galaxies are often embedded in large envelopes of HI (e.g., Begum et al. 2008) which makes it unlikely that the Lyman continuum photons find their way completely out of a galaxy and into the IGM.  Moreover, observations which have attempted to directly detect escaping Lyman continuum photons have all found upper-limits $\lesssim$10\%  (e.g., Leitherer et al. 1995, Bergvall et al. 2005, Grimes et al. 2007, Siana et al. 2007).  These detection experiments have been performed on starbursting galaxies where leakage is thought to most likely occur.  In contrast, the majority of the dwarfs in the sample are relatively normal in their current star formation activity (Lee et al. 2009a).  It thus seems unlikely that leakage can be the cause for the trend.

Ionizing photons can also be ``lost" in a second way,
via absorption of Lyman continuum photons by dust (i.e.,
dust absorption that Balmer decrement based
attenution corrections cannot account for since the
ionizing photons encounter dust before they
are able to ionize hydrogen).  However, it seems 
even more unlikely that this effect can cause the 
observed trend in the H$\alpha$-to-UV ratio since
the absorption would need to be larger in relatively transparent 
dwarf galaxies than more luminous, dustier systems.\footnote{
Further discussion on the direct absorption of Lyman continuum 
photons by dust in more luminous systems can be found in 
Buat et al. (2002), Hirashita et a. (2003) and 
Iglesias-Paramo et al. (2004).  Such absorption is 
unlikely to be a salient issue in understanding the 
systematic decline in the H$\alpha$-to-UV ratio 
so we do not address it in any detail here.  
We do note however that the fraction of ionizing
photons absorbed by dust cannot be very large, and is 
probably less than $\sim$10-15\%, based on the relative
agreement of dust corrected UV and H$\alpha$ SFRs at the
luminous end of our sample.}

\subsection{Starbursts in Dwarf Galaxies}\label{sec:bursts}

Previous work tentatively attributed possible trends in the H$\alpha$-to-FUV ratio to a systematic increase in the prevalence of bursts in the recent star formation histories of low mass galaxies (e.g., Bell \& Kennicutt 2001).  The fiducial expected ratio is an ``equilibrium'' value which is calculated assuming that the SFR has been constant over a time period that is long enough for the birth of stars responsible for the FUV and H$\alpha$ emission to balance their deaths.  Variations in the SFR over timescales $\sim$100 Myr would disrupt this equilibrium.  In the time following a burst of star formation, a deficiency of ionizing stars develops as they expire relative to lower-mass, longer-lived stars that also contribute to the UV emission, and the H$\alpha$-to-FUV flux ratio will thus be lower than expected (e.g., Sullivan et al. 2000, Sullivan et al. 2004, Iglesias-Paramo et al. 2004).

To examine whether variations in the SFR can account for the decline of the H$\alpha$-to-FUV ratio of the magnitude that is observed, synthesis modeling is needed.  Here the work of Iglesias-Paramo et al. (2004) can be directly applied.  Iglesias-Paramo et al. (2004) sought to use the H$\alpha$-to-FUV flux ratio to constrain the effects of the cluster environment on the recent SFH of spiral galaxies, and computed a grid of Starburst99 models for a range of burst amplitudes and durations, assuming a Salpeter IMF.  The bursts were superimposed on 13 Gyr-old models which were evolved assuming an exponentially declining SFR, with decay timescales typical of values normally used to fit star forming galaxies (from 3 to 15 Gyr).  Variations in the predicted H$\alpha$-to-FUV flux ratio were found to be insensitive to differences in decay timescale values within this range.  The results of their modeling are tabulated, and it is shown that in order to produce H$\alpha$-to-FUV flux ratios that are on average about a factor of two lower than the fiducial equilibrium value in a given sample, bursts which elevate the SFR of the galaxies by a factor of 100 for a 100 Myr duration are required.  However, such large amplitude bursts do not appear to commonly occur in the overall dwarf galaxy population.  The most direct constraints on burst amplitudes and durations in dwarf galaxies are provided by studies which reconstruct star formation histories from resolved observations of stellar populations in the nearest low-mass systems.  Most recently, Weisz et al. (2008) and McQuinn et al. (2009) have analyzed HST/ACS imaging of dwarf galaxies in the M81 group and a small sample of low-mass starburst systems, respectively.  Typical burst amplitudes range from a few to $\sim$10, an order of magnitude smaller than the factor of 100 bursts required to depress the H$\alpha$-to-FUV flux ratios by a factor of 2.  Further, the 11HUGS sample itself provides a statistical constraint on the average dwarf galaxy starburst amplitude via the ratio of fraction of star formation (as traced by H$\alpha$) concentrated in starbursts to their number fraction (Lee et al. 2009a).  The high degree of statistical completeness of 11HUGS makes this calculation possible for galaxies with $M_B<-15$, and again, the burst amplitude is found to be relatively modest ($\sim$4), at least for this population of dwarfs.
From the above arguments it appears that modest starbursts by themselves, even if they are 
common in low-mass systems, cannot explain the observed deviation in 
the H$\alpha$-to-FUV flux ratios from the expected value.

\subsection{Stochasticity in High Mass Star Formation at Low SFRs}\label{sec:stoch}
An implicit assumption in all stellar population synthesis models (and the resulting calibrations based upon them) is that statistical numbers of stars spanning a full range of masses, typically to a limit of M$_{\ast}$= 100 M$_{\odot}$, are produced.  In the regime of ultra-low SFRs where only a handful of O-stars are formed in a given system over timescales comparable to their lifetimes (i.e., a few million years), this assumption is clearly violated.  Under these circumstances, H$\alpha$ emission can appear deficient, or even absent, although lower mass star formation, which dominates the total mass formed, may be taking place and the underlying IMF {\it is invariant}.  Such statistical effects on the H$\alpha$ luminosity has been previously explored for example by Boissier et al. (2007) and Thilker et al. (2007) in the context of extended UV disks.  More detailed work has been done by Cervi\~{n}o \& Valls-Gabaud (2003), Cervi\~{n}o et al. (2003) and Cervi\~{n}o \& Luridiana (2004) for modeling properties of low mass stellar clusters and other simple, single-age stellar populations.  To evaluate whether such effects can account for our observations, we first perform a simple back-of-the-envelope calculation, and then examine results from the Monte Carlo simulations of Tremonti et al. (2009).  

We first estimate when Poisson fluctuations should begin to affect the H$\alpha$ output by computing the SFR at which the number of O-stars on the main sequence at any given time will be less than 10.  Assuming a Salpeter IMF with standard mass limits of 0.1-100\msunyr, and considering that only stars with masses $>$18 M$_{\odot}$ significantly contribute to the ionizing flux (e.g., Hunter \& Elmegreen 2004, appendix B), simple integrals show that there are 10 ionizing stars for a total stellar mass of 4.3$\times$10$^{3}$ M$_{\odot}$ formed.  To compute the corresponding SFR above which the H$\alpha$ flux should be robust against stochasticity in the formation of ionizing stars, we divide this mass by the 3 Myr lifetime of a 100 M$_{\odot}$ star (Schaerer et al. 1993), which yields 1.4$\times$10$^{-3}$ \msunyr (log SFR=-2.8).  This is a conservative limit compared to calculations based on the formation of single O-stars (e.g., Lee et al. 2009a; Meurer et al. 2009)\footnote{In Lee et al. (2009a), the calculation at the end of \S2.1.1 was intended to estimate the SFR required to produce a single O-star at any given time for a Salpeter IMF with mass limits of 0.1 and 100 M$_{\odot}$, and not 10 O-stars as stated there.}.  However, Figures \ref{fig:uvhacomp} and \ref{fig:uvharesidualdustcor} show that systematic deficiencies in the H$\alpha$-to-FUV ratio begin at SFRs that are almost an order of magnitude higher (log SFR$\sim$-2.0).   At log SFR=-2.8 the ratio is not just beginning to deviate from the expected value; it is already more than a factor of two lower.  This back-of-the envelope calculation suggests that stochasticity in the sampling of an invariant Salpeter IMF alone is unlikely to give rise to the observed trends.  The same holds true for the Kroupa (2001) and Chabrier IMFs (2003), for which the computed SFR limit would be 1.5 times lower.  In all cases the FUV flux is negligibly affected.

\begin{figure}
\epsscale{0.65}
\plotone{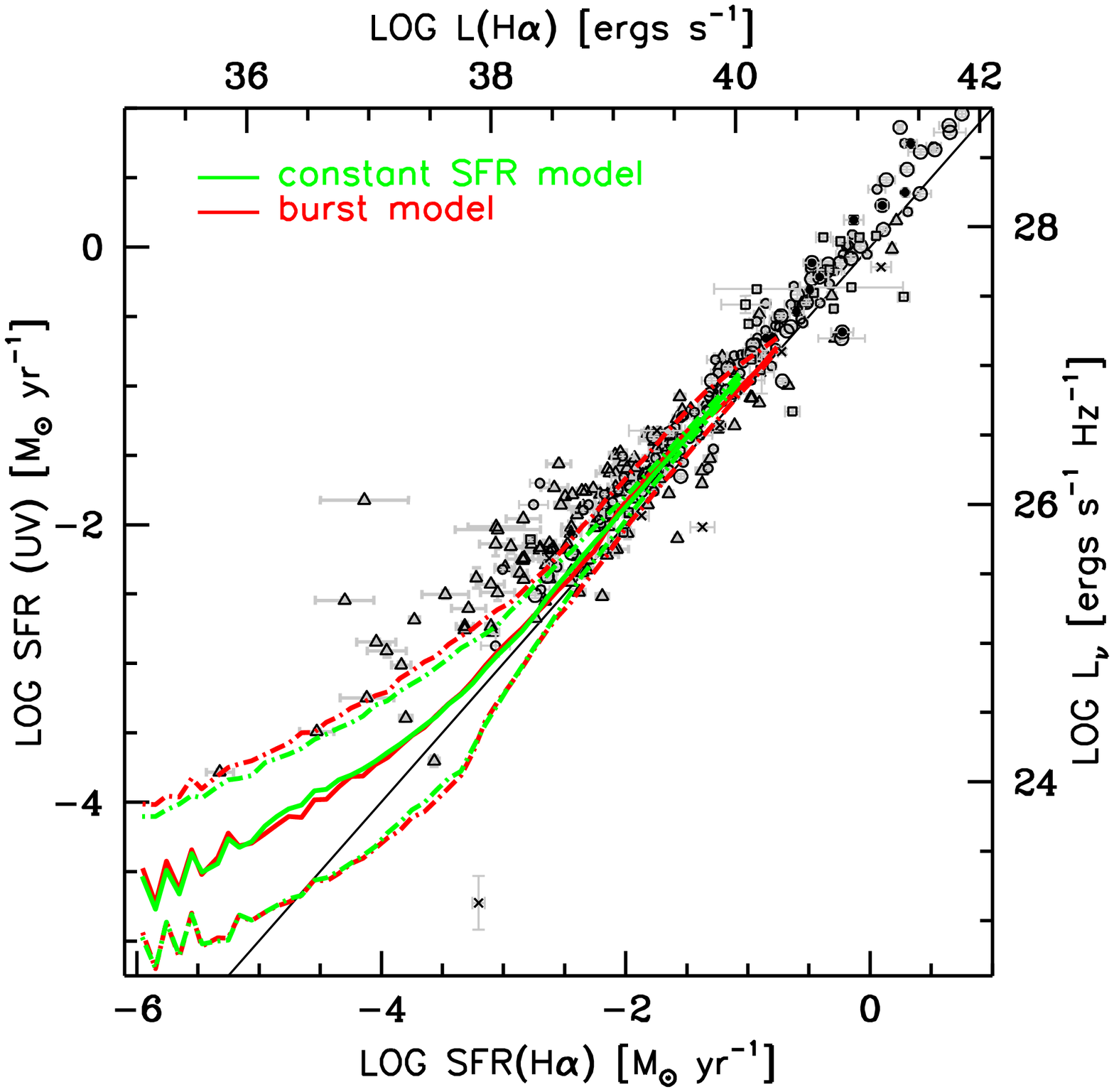}
\epsscale{0.9}
\plotone{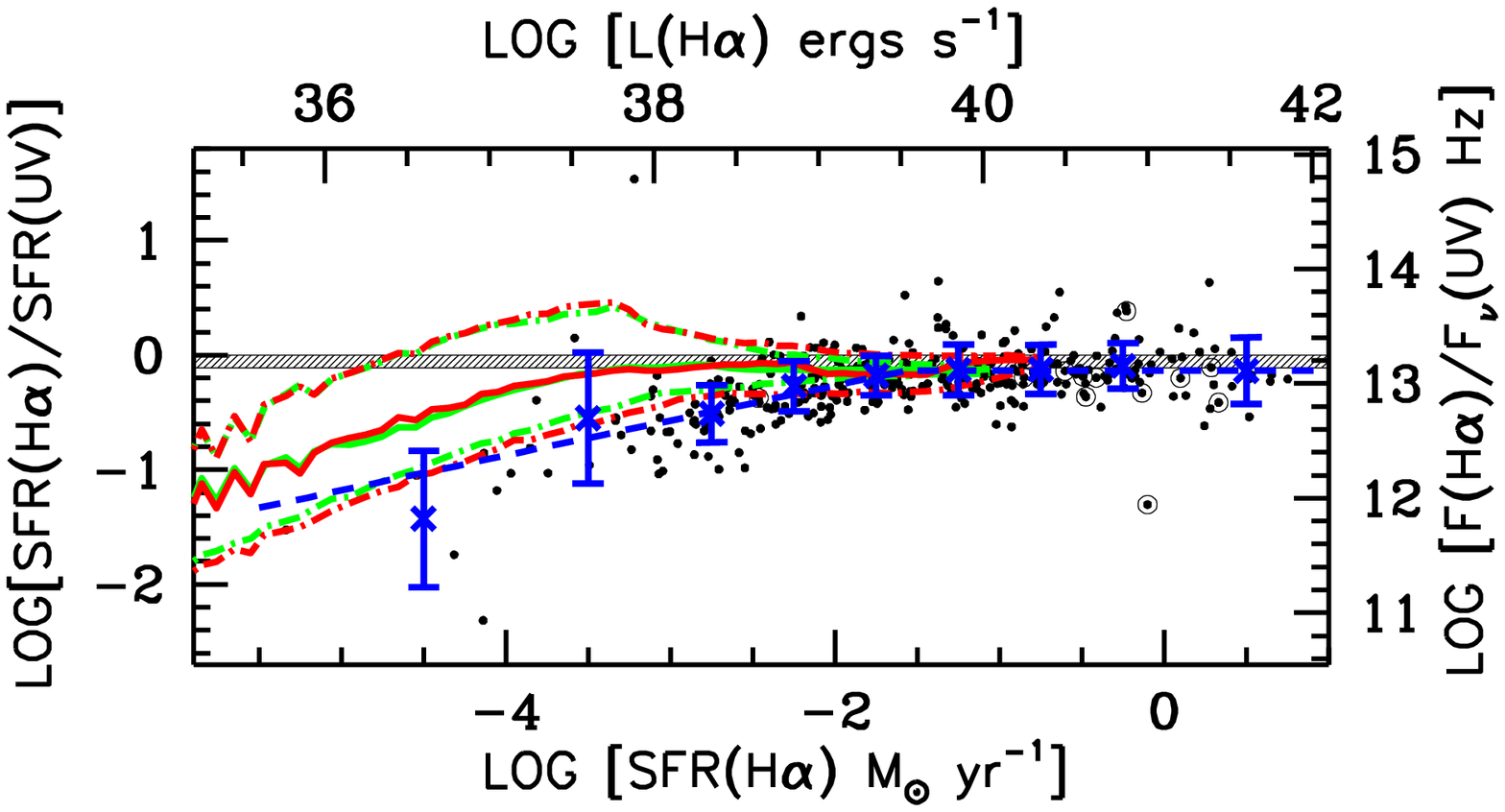}
\vspace*{-2.8in}
\caption{Top panel: Same as Figure~\ref{fig:uvhacomp}, but corrected for attenuation by dust internal to the galaxies as described in Section~\ref{sec:dustcorr}, and with predictions from the stochastic models of Tremonti et al. (2007; 2009) overplotted.  Bottom panel: Same as the top panel of Figure~\ref{fig:uvharesidualdustcor}, but again with the stochastic model predictions overplotted.  The median predicted values (solid line) as well as the values at the 2.5 and 97.5 percentile points (dotted lines) are shown.  The green curves represent a model where the star formation rate is constant at all times, while the red curves show a model where bursts with an amplitude of four and a duration of 100 Myrs are added every 500 Myrs.}\label{fig:stoch}

\end{figure}

Tremonti et al. (2009) have preformed detailed Monte Carlo simulations to confirm these conclusions. Such calculations not only allow for a more robust prediction of the mean deviations due to
stochasticity in the production of massive ionizing stars, but also provide quantitative predictions of the
accompanying increase in scatter.  In brief, we generate a pool of stars which have a mass distribution
given by the Salpeter IMF, and assign each star a mass-dependent main
sequence lifetime, ionizing photon production rate, and FUV luminosity.  We randomly draw stars from the
pool at a rate proportional to the desired SFR and follow the ensemble population over time.  SFRs are 
calculated from the FUV and H$\alpha$ luminosities at t$>$1 Gyr.  We have run simulations for a large
grid of SFRs; the results are overplotted on the data in Figure~\ref{fig:stoch}.  The median predicted values (solid line) as well as the values at the 2.5 and 97.5 percentile points (dotted lines) are shown.  The median flux ratio begins to significantly deviate from the fiducial value only for log SFR$< -3$, which is consistent with expectations based on the back-of-the-envelope estimate given above.  

We note that the results of our calculations agree with the work of Cervi\~{n}o et al. (2003) and Cervi\~{n}o \& Luridiana (2004), who have used statistical formalism to compute values for the ``Lowest Luminosity Limit'' (LLL), the limiting stellar mass formed in a single event whose properties can be modeled without accounting for stochastic effects.  According to their calculations, $\sim$3$\times$10$^{3}$ M$_{\odot}$ is the mass associated with the LLL in the production of Lyman continuum photons.  Again, taking the O-star lifetime to be $\sim$3 Myr as above, this limiting mass would correspond to log SFR=-3.  However, as Cervi\~{n}o et al. (2003) discuss, statistical effects may be present in formation events involving up to ten times the LLL mass.  Based upon the Monte Carlo simulations, these effects manifest as an increase in scatter in the H$\alpha$-to-FUV ratio beginning at log SFR$\sim$-2, although the median value of the ratio remains stable until log SFR$\sim$-3.
 
Therefore, to summarize, although stochasticity in the sampling of an invariant IMF does result in a declining H$\alpha$-to-FUV ratio with SFR, we find that the mean effect is not large enough to account for the observations.  One important caveat regarding this conclusion however is that our calculations assume that the SFR is constant.  As discussed in Section~\ref{sec:bursts}, moderate variations of the SFR with time are seen in the recent star formation histories reconstructed from resolved stellar population data of nearby dwarf galaxies (e.g., Dohm-Palmer et al. 1998; Weisz et al. 2008), and stochasticity may amplify the effect of such variations on the H$\alpha$-to-FUV ratio.  As illustrated in Figure~\ref{fig:stoch} however, an initial analysis where bursts with 100 Myr duration and factor of four amplitude are included in our Monte Carlo simulations show that this is unlikely to produce significant changes in the predictions based on a constant SFR alone.  However, more comprehensive modeling covering a range of star formation histories is needed, and these issues are further examined in Tremonti et al. (2009).  In particular, histories where the SFR shows short timescale ($\sim$10 Myr) fluctuations of factors of 10 or greater may be considered.  Such histories may be characteristic of the most extreme dwarfs in our sample, in which only one or two HII regions are visible, and where such fluctuations are not necessarily indicative of starburst activity, but rather of Poisson noise in the formation of individual star clusters.  

\subsection{IMF}

Finally, we examine the basal assumption required for the calibration of all star formation indicators, the form of the IMF.  Naturally, indicators which trace highly luminous but relatively rarer massive stars are particularly sensitive to its form.   However, the IMF is generally assumed to be invariant, and if so, the H$\alpha$-to-FUV flux ratio should be constant, modulo the other factors discussed above.  
Uncertainties in a {\it universal} IMF therefore cannot themselves produce trends in the H$\alpha$-to-FUV flux ratio.  Rather, systematic variations which result in a deficiency of ionizing stars in dwarf galaxies would be needed to explain the observations.   This possibility must be considered, given that we cannot explain the magnitude of the H$\alpha$-to-FUV offset with the other parameters that have been explored thus far.

Evaluating the plausibility of this scenario in present day galaxies has been an issue of much recent work.
For example, Hoversten \& Glazebrook (2008; HG08) examined the H$\alpha$ EW (equivalent width) and optical colors of a large sample of galaxies taken from the Sloan Digital Sky Survey (SDSS).  They find that the H$\alpha$ EWs are systematically low for the optical colors in low-luminosity galaxies given expectations based on continuous star formation.  This is consistent with our results for a more limited sample drawn from the Local Volume (Lee 2006, 2008).  However, whereas Lee (2006, 2008) concluded that such discrepancies could be explained by invoking starbursts in the recent past histories of dwarfs (most systems would then be in a post-burst state where the ionizing stars from the burst have died off, but the intermediate mass stars continue to contribute to the continuum flux density at H$\alpha$, thus depressing the H$\alpha$ EW), the analysis of HG08 appears to have ruled this out.  The much larger SDSS sample of HG08 enabled more detailed modeling which probed the temporal variation of the EWs and colors over a burst cycle.  HG08 found that fine-tuning to coordinate burst times was necessary to reproduce the data, and that a more likely explanation was that low-luminosity galaxies have an IMF deficient in massive stars.

Recently, Meurer et al. (2009) have examined correlations in the H$\alpha$-to-FUV flux ratio with the H$\alpha$ and $R$-band surface brightnesses, using the HI-selected samples of the Survey of Ionization in Neutral Gas Galaxies (SINGG; Meurer et al. 2006) and the Survey of UV emission in Neutral Gas Galaxies (SUNGG).  Meurer et al. find that the H$\alpha$-to-FUV flux ratio decreases with decreasing surface brightness.  After ruling out the more likely causes of the systematic in an analysis parallel to the one presented here, they have concluded that it is evidence for a non-uniform IMF.  Their study is complementary to ours since they have examined correlations in the H$\alpha$-to-FUV flux ratio with the H$\alpha$ and $R$-band surface brightness instead of integrated luminosities, and use an independent dataset.  The galaxy sample however is weighted more heavily toward more luminous star-forming systems (e.g. Paper I, Figure 7 \& 8); there are only $\sim$10 galaxies with dust corrected SFR(H$\alpha$)$<$0.01 \msunyr and this precludes a robust determination of the observational trend with integrated luminosity as presented here (M. Seibert, private communication).

On the theoretical side, Kroupa \& Weidner (2003) and Weidner \& Kroupa (2005, 2006) have formulated a model based on both statistical arguments and observational constraints which produces IMFs that appear to be steeper for galaxies with lower SFRs.  Several critical assumptions are made.  The first is that all stars are born in stellar clusters, which universally form stars according to the Kroupa (2001) IMF.  The formation of stellar clusters themselves is also governed by a power law.  While the convolution of cluster and stellar IMFs by itself does not produce effects beyond simple random sampling of the stellar IMF alone, the model also assumes that ({\it i}) the maximum mass of a cluster formed in a given galaxy, $m^{max}_{cl}$, depends on its SFR, and ({\it ii}) the maximum mass of a star formed in a given cluster, $m^{max}_{\ast}$, depends on the total cluster mass $M_{cl}$, and is lower than the strict limit where $m^{max}_{\ast}=M_{cl}$.  The claim is that both $m^{max}_{cl}$(SFR) and $m^{max}_{\ast}(M_{cl})$ are both robustly constrained by observations.  Further, it is shown that $m^{max}_{\ast}(M_{cl})$ is also well-modeled by the statistical condition:

\begin{equation}
\int_{m^{max}_{\ast}}^{150\;M_{\odot}}\xi(m)dm=1 \;\; \bigwedge \;\;
M_{cl}=\int_{m^{min}_{\ast}}^{m^{max}_{\ast}}m\xi(m)dm
\end{equation}

\noindent All these effects taken together can be expressed as:

\begin{equation}
\xi_{IGIMF}(m,t)=\int^{m^{max}_{cl}(SFR(t))}_{m^{min}_{cl}}\xi_{\ast}(m\leq m^{max}_{\ast}(M_{cl})) \;\;\xi_{cl}(M_{cl})\;\;dM_{cl}
\end{equation} 

\noindent where $\xi_{IGIMF}$ is referred to as the ``Integrated Galaxial IMF.''

\begin{figure}
\epsscale{0.65}
\plotone{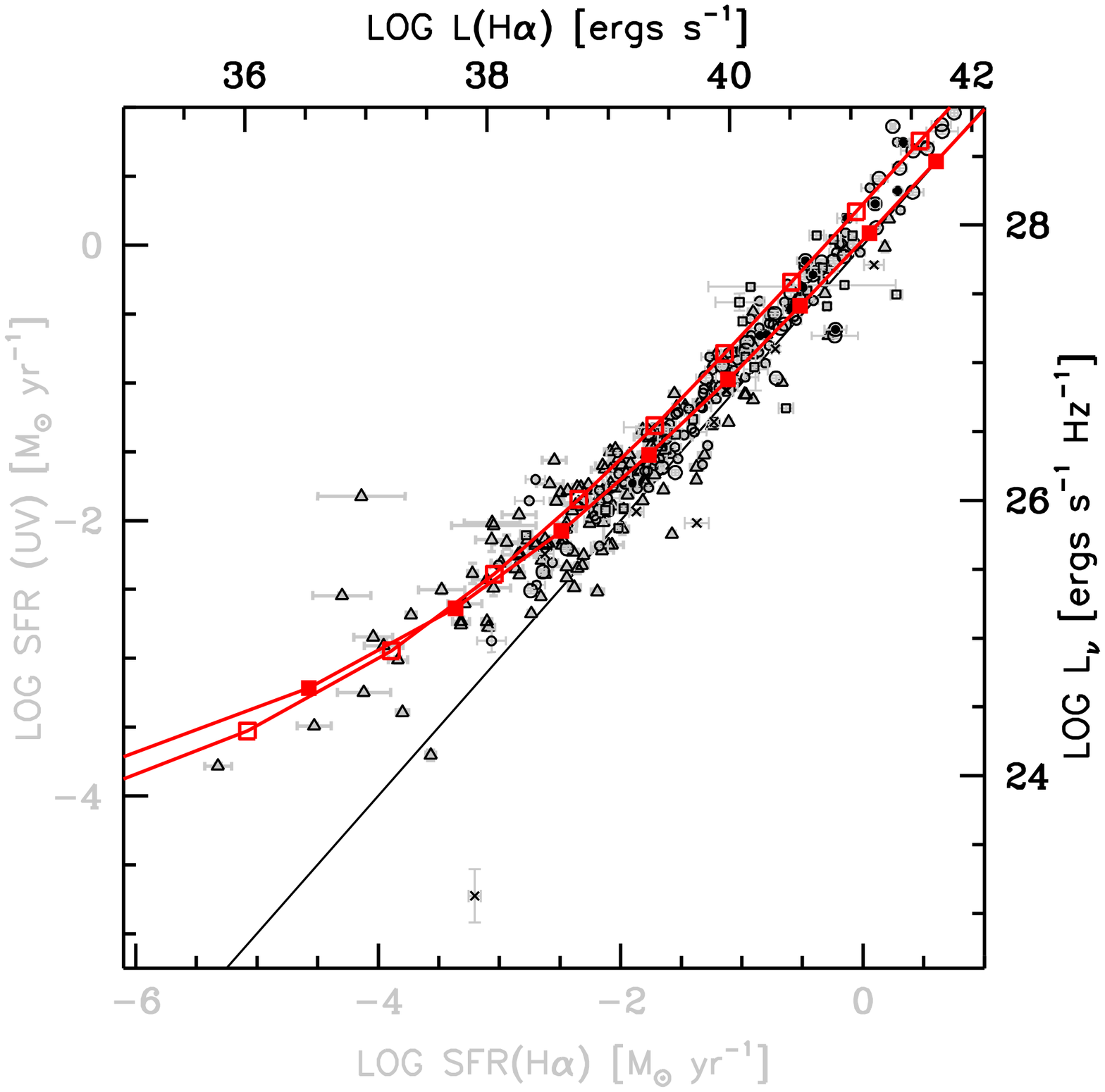}
\epsscale{0.9}
\plotone{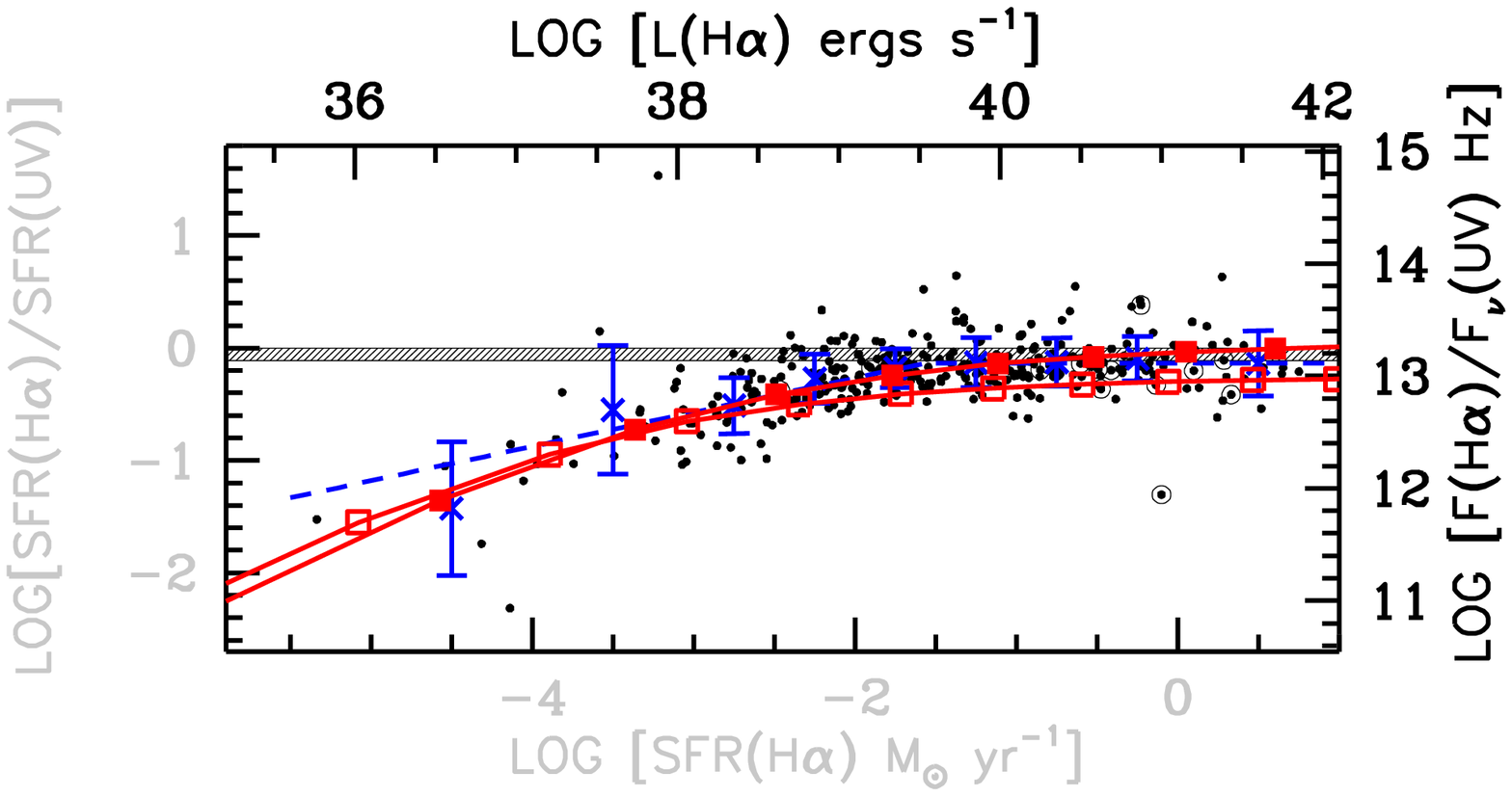}
\vspace*{-3in}
\caption{Top panel: Same as Figure~\ref{fig:uvhacomp}, but corrected for attenuation by dust internal to the galaxies as described in Section~\ref{sec:dustcorr}.  Predictions based on the IGIMF model (Kroupa \& Weidner 2003; Pflamm-Altenburg, Weidner \& Kroupa 2009) are overplotted.  Bottom panel: Same as the top panel of Figure~\ref{fig:uvharesidualdustcor}, but again with predictions based on the IGIMF model overplotted. The open and filled red squares represent the IGIMF ``standard'' and ``minimal1'' models, respectively.  The bottom and left hand axes are shown in gray to signify that the K98 SFR scales would not be valid at low SFRs under the assumptions of the IGIMF model.}\label{fig:igimf}
\end{figure}

Recently, Pflamm-Altenburg, Weidner \& Kroupa (2007; 2009) have used this model to generate predictions for trends in FUV and H$\alpha$ properties.  They consider the case where the cluster mass function  $\xi_{cl}$ has a Salpeter slope (referred to as the ``standard model'') and also examine the consequences of assuming one which has a higher ratio of high-mass to low-mass clusters, i.e., $\alpha_{cl}$=1.0 for $5 M_{\odot} \leq M_{cl} < 50 M_{\odot}$ and $\alpha_{cl}$=2.0 for $50 M_{\odot} \leq M_{cl} \leq m^{max}_{cl}(SFR)$.  Adopting this $\xi_{cl}$ results in smaller relative deviations from a universal Kroupa IMF and is referred to as the ``minimal1'' model.  The predictions are overplotted on the data in Figure~\ref{fig:igimf}, where the open red squares represent the standard model, while the filled red squares represent the minimal1 model.  Both are in good agreement with the data at the faint end, while the data for the more luminous galaxies in the sample prefer the minimal1 model.  Therefore, the IGIMF model is able to account for the declining H$\alpha$-to-FUV ratio observed in our Local Volume galaxy sample. 

A next critical test of the IGIMF model would involve computing the expected scatter in the relationship given plausible star formation histories, since it is not clear whether the IGIMF will allow for H$\alpha$-to-UV ratios higher than the predicted curve (as seen in the data, modulo extinction), given the deterministic upper-mass limits on the IMF in the low activity regime.  Monte Carlo simulations analogous to the ones described in the previous section should also be run with the IGIMF as the input IMF to check that the combination of stochasticity, variations in the SFH and the IGIMF do not lead to {\it larger} declines in the H$\alpha$-to-UV ratio than is seen in the data.
Finally, further work is still also required to understand the relationship between the IGIMF model and the physics of the fragmentation and collapse of gas into stars. 

\subsection{L(H$\alpha$) as an SFR Indicator at Low Activities}

\begin{figure}
\plottwo{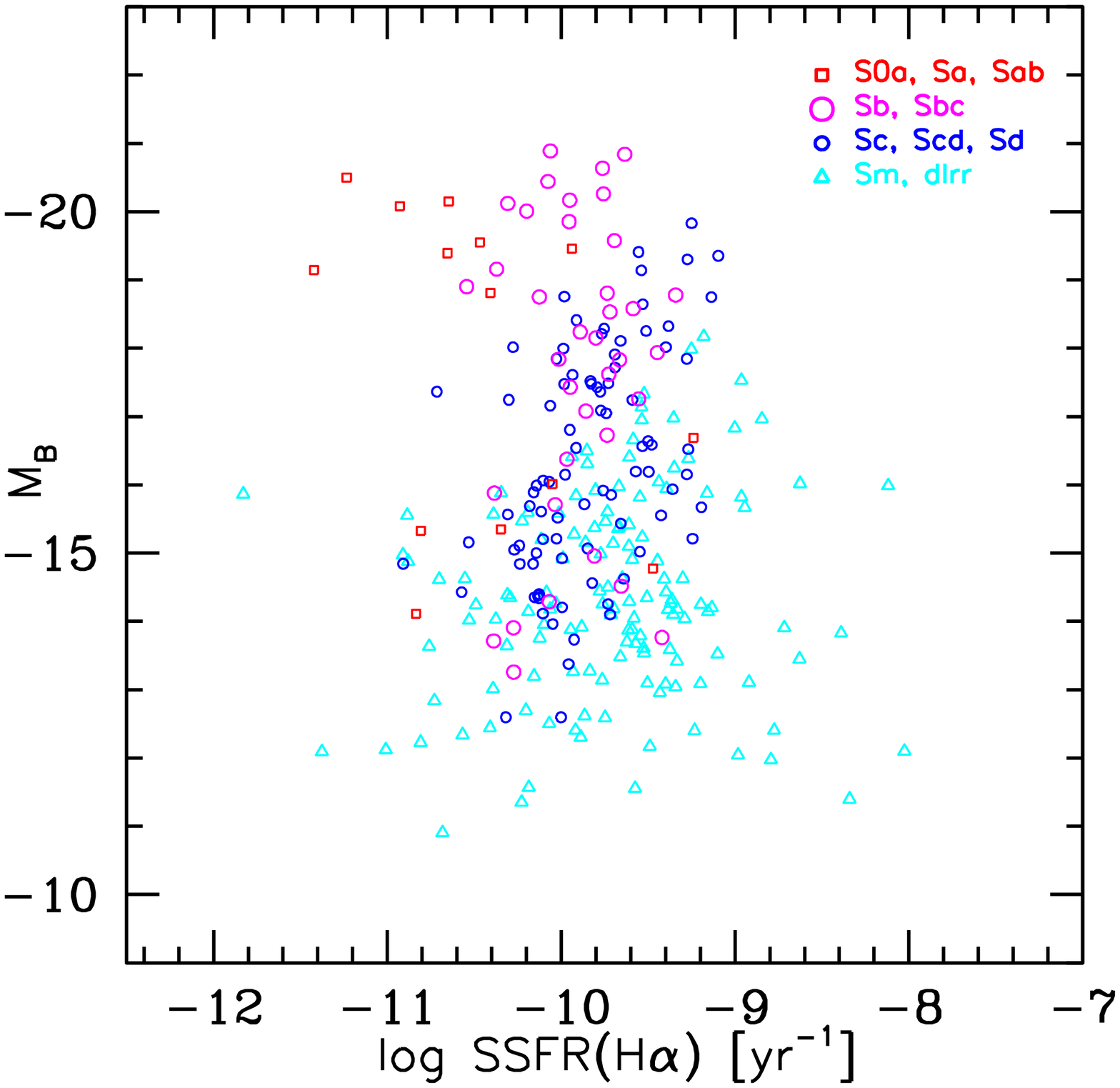}{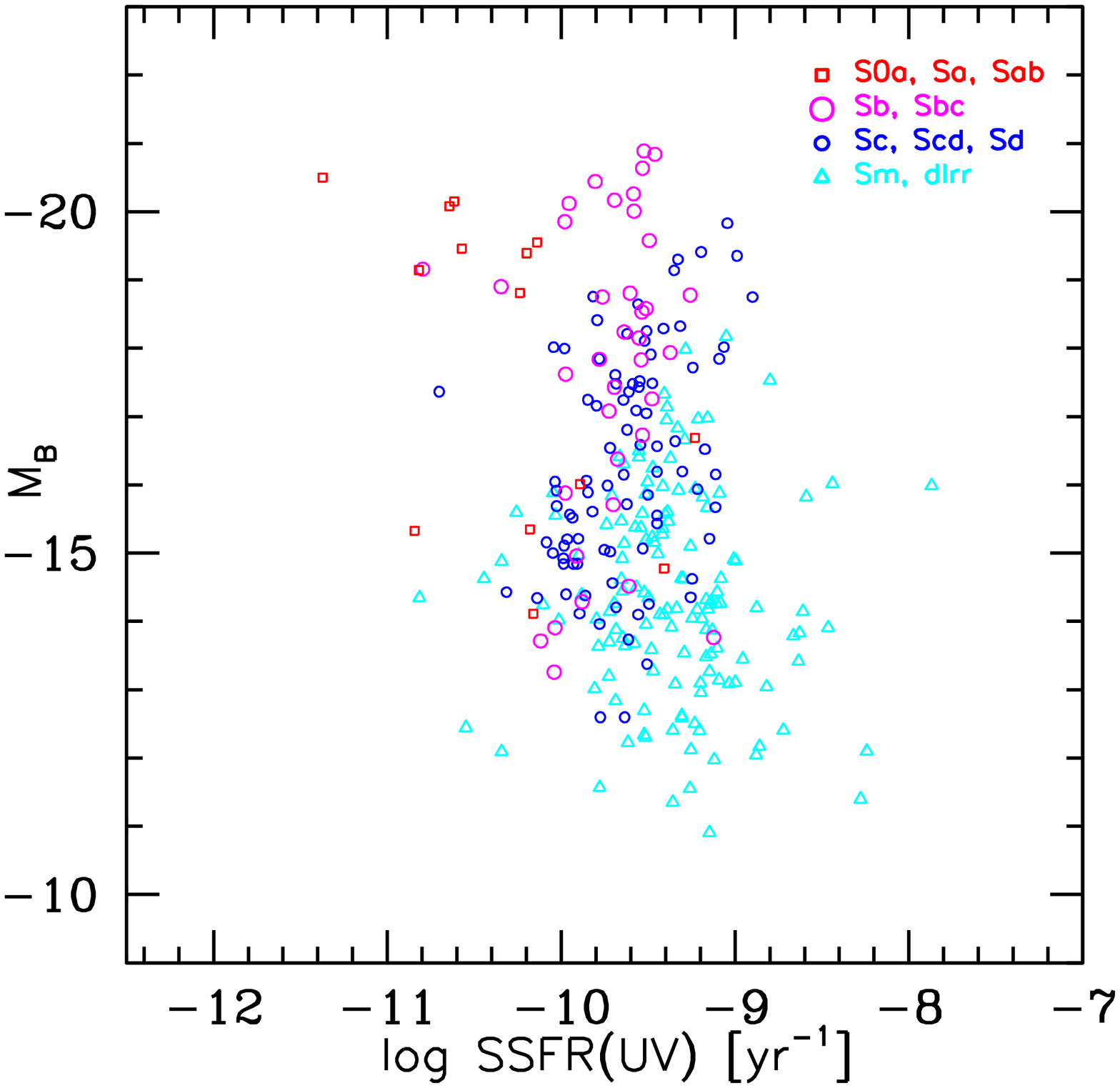}
\caption{The variation of the specific SFR (SSFR; the SFR per unit stellar mass) with $M_B$ where the SSFR has been dust corrected and is based on the H$\alpha$ (left panel) and FUV luminosities (right panel).  These plots are analogous to those presented in Lee et al. (2007) which directly used the H$\alpha$ EW to trace the SSFR.  In that paper, a transition at $M_B\sim-15$, characterized by an increase in scatter toward low luminosities, was identified.  The same structure is apparent in these figures, but with the notable distinction that the relative increase in scatter is reduced when the SSFR is calculated from the FUV luminosity.}\label{fig:ssfr}
\end{figure}

Whether or not IMF variations (or other unidentified issues) are the true culprits underlying the systematic variation in the H$\alpha$-to-FUV ratio, it seems reasonable to surmise that the FUV luminosity is the more robust SFR indicator in individual galaxies with low total SFRs and low dust attenuations.  This is also supported by the finding that the increase in scatter at low galaxy luminosities in the H$\alpha$ equivalent width (which traces the SFR per unit stellar mass), as discussed in Lee et al. (2007), is mitigated when SFR per unit mass is instead calculated from the FUV.  Figure~\ref{fig:ssfr} illustrates the variation of the specific SFR (SSFR; the SFR per unit stellar mass) with $M_B$ where the SSFR has been dust corrected and is based on the H$\alpha$ (left panel) and FUV luminosities (right panel).  The SFRs and attenuations corrections are identical to those used throughout this paper, while the stellar masses are based upon $M_B$ and color-dependent mass-to-light ratios as described in Bothwell et al. (2009).  These plots are analogous to those presented in Lee et al. (2007).  The previously reported increase in scatter near $M_B\sim-15$ is apparent in both panels.  However, whereas the scatter in SSFR(H$\alpha$) increases toward the lowest luminosities by $\sim$60\%, the increase is only $\sim$25\% for SSFR(FUV).  Given the longer lifetimes and greater numbers of stars which dominate the UV output of a galaxy, this result is reasonable as the UV will be less affected by purely stochastic variations in the formation of high mass stars.  This relative reduction in the increase in scatter is also generally consistent with the predictions of our Monte Carlo simulations for SFR$\lesssim$0.01 \msunyr\ as reported in Tremonti et al. (2009) and discussed in Section~\ref{sec:stoch}.

Therefore, the results of our analysis suggest that when possible, 
the SFR should be based on the FUV luminosity when SFR$\lesssim$0.01 \msunyr.  {\it H$\alpha$ is a less reliable indicator when the integrated, galaxy-wide log $L(H\alpha)$ [ergs s$^{-1}$]$\lesssim39.4$.}  However, for those instances where only an H$\alpha$ measurement is available, we provide an empirical re-calibration of SFR(H$\alpha$), using the extinction corrected K98 SFR(FUV) as the reference value.  For an observed, non-dust corrected, integrated L(H$\alpha$)$<2.5 \times 10^{39}$ ergs s$^{-1}$:
 
\begin{equation}\label{eq:calib}
\mbox{log (SFR}\;\;[\mbox{M}_{\odot} \mbox{yr}^{-1}]) = 0.62\; \mbox{log}(7.9 \times 10^{-42} L(H\alpha) [\mbox{ergs s}^{-1}])-0.47.
\end{equation}

\noindent This relation is the result of a linear least squares fit of SFR(FUV), where L(FUV) has been corrected for internal dust extinction, as a function of SFR(H$\alpha$), which has {\it not} been dust corrected.
Thus, 
the SFR calculated using this expression will represent intrinsic, dust unattenuated values as appropriate for the dust content typical of local galaxies, trace the activity averaged over the timescale of the FUV emission ($\sim$100 Myrs) instead of the native H$\alpha$ instantaneous timescale ($\sim$3 Myrs), and carries the same assumptions (fiducial stellar models, Salpeter IMF) as the K98 UV SFR calibration.  The random error on the SFR, which will increase with decreasing $L(H\alpha)$, can be estimated from the 1-$\sigma$ scatter in SFR(H$\alpha$)/SFR(FUV) as listed in Table 2.  

\begin{figure}
\epsscale{0.7}
\plotone{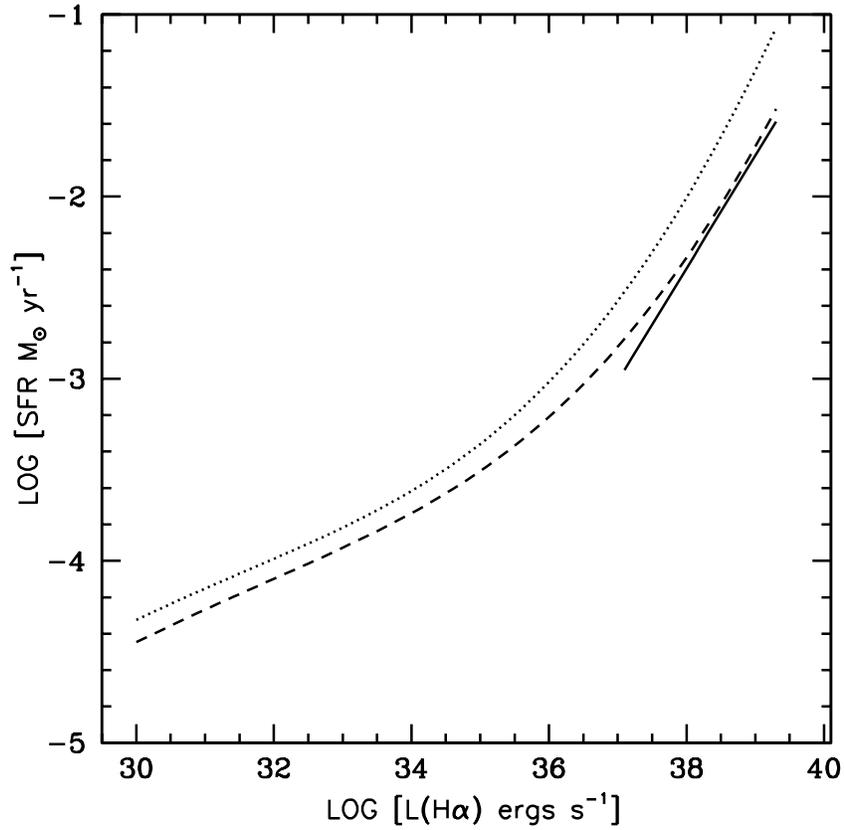}
\caption{The empirical calibration of L(H$\alpha$) as a SFR indicator in the regime of low integrated activities based on the 11HUGS dataset (equation~\ref{eq:calib}; solid line), compared with the Pflamm-Altenburg, Weidner \& Kroupa (2007) predictions for the IGIMF model.  The dotted and dashed curves represent the IGIMF ``standard'' and ``minimal1'' models, respectively.}\label{fig:calib}
\end{figure}

As with any SFR prescription, it is important to be aware of the ranges of applicability of Equation~\ref{eq:calib} and its limitations.  The relation has been calibrated with integrated SFRs of local dwarf irregular galaxies for which log L(H$\alpha$)$\gtrsim$37, so application to other systems with low total SFRs (e.g., massive elliptical galaxies, or galaxies with log L(H$\alpha$) [ergs s$^{-1}$]$<$37) may not be warranted.\footnote{Note that although the dataset extends to log L(H$\alpha$) [ergs s$^{-1}$]$\sim$36, there are only 5 galaxies with log L(H$\alpha$) [ergs s$^{-1}$]$<$37.  Clearly, a larger sample is needed to constrain systematics in the H$\alpha$-to-FUV ratio and the empirical re-calibration in this regime.}  We also note that higher order terms may be required to more accurately describe the systematic underestimation of the SFR in this regime, although the current dataset does not adequately constrain such terms.  For example, Pflamm-Altenburg et al. (2007) offer a calibration based on the IGIMF model that is described by a 5th order polynomial.  In Figure~\ref{fig:calib}, our empirical relation is compared with the IGIMF prescriptions.  The IGIMF ``minimal1'' model is consistent for log L(H$\alpha$)$\gtrsim$38, but predicts up to 0.2 dex more star formation in the regime where there is sufficient data available to calibrate our relation.

\section{Summary and Conclusions}

Using H$\alpha$ flux measurements and GALEX UV photometry from the 11HUGS Legacy program (Kennicutt et al. 2008; Lee et al. 2009b),  we evaluate the consistency between star formation rates (SFRs) inferred from H$\alpha$ and the far UV (FUV) continuum over 5 orders of magnitude, down to ultra-low SFRs of $\sim$0.0001 \msunyr, assuming standard conversion recipes in which the SFR scales linearly with luminosity at a given wavelength.  The primary observational result is that the H$\alpha$ luminosity underestimates the SFR relative to the FUV luminosity in dwarf galaxies that are roughly less active than the Small Magellanic Cloud; that is, the observed H$\alpha$-to-FUV flux ratio systematically decreases with declining luminosity.  The effects of uncertainties in the stellar evolution tracks and model atmospheres, non-solar metallicities, non-constant SFHs, possible leakage of ionizing photons/departures from Case B recombination, dust attenuation, stochasticity in the formation of high-mass stars, are all considered.  However, none of these potential drivers acting alone are able to explain the magnitude of the observed systematic.  The underlying cause for the trend is not clear, although we cannot rule out variations in the IMF which result in a deficiency of high mass, ionizing stars in dwarf galaxies.  In fact, the semi-empirical IGIMF model of Kroupa \& Weidner (2003), which results in such a deficiency, predicts a trend in the H$\alpha$-to-FUV flux ratio that is strikingly consistent with our data.  It is also possible that some combination of effects may conspire to produce the observed systematic.  

More work is needed to further elucidate the nature of the systematic as presented in this paper and in Meurer et al. (2009).  For example, stochasticity may amplify the effects of bursty or non-uniform ``flickering'' star formation histories on the H$\alpha$-to-FUV ratio, as discussed in Section~\ref{sec:stoch} and we investigate such effects in greater detail in Tremonti et al. (2009).  Another obvious issue is the difficulty in evaluating whether the perceived inconsistency is simply due to unknowns in stellar evolution modeling of massive stars at low metallicity.  Also, although it seems unlikely that the leakage of ionizing photons can be significant enough to account for the low H$\alpha$-to-FUV flux ratio observed, it remains a possibility that photons can preferentially escape in a direction perpendicular to the disk in low mass galaxies.  Tracing the formation of high mass, ionizing stars in an independent way, such as with radio continuum measurements, may be useful.  Examining the relative spatial distributions of emission within UV-bright, H$\alpha$-faint galaxies may also provide clues on the origin of the systematic.  In a number of dwarfs, we observe that whereas the FUV emission extends over the optical disk, the H$\alpha$ (and infrared emission) is far more localized in a few scattered clumps.  
Finally, to further test the possibility of systematic variations in the IMF, other consequences must be followed up to ensure that at a minimum, these yield a consistent picture.  As pointed out by K\"{o}ppen et al. (2007), IMFs deficient in high mass stars should imprint signatures in the observed metal abundance ratios, and these must be carefully reexamined in this context.

\acknowledgments
JCL thanks the astronomers at Carnegie Observatories, in particular Alan Dressler, Luis Ho, Andy McWilliam, and Steve Schechtman, for helpful conversations over the course of this work.
Support for JCL has been provided by NASA through Hubble Fellowship
grant HST-HF-01198 awarded by the Space Telescope Science
Institute, which is operated by the Association of Universities for
Research in Astronomy, Inc., for NASA, under contract NAS5-26555.
AGdP acknowledges partial support from the Spanish Ram\'{o}n y Cajal program and
the Programa Nacional de Astron\'{o}mica y Astrof\'{i}sica under grant AYA
2006-02358.
CAT thanks the Alexander von Humboldt Foundation for their generous support.
This research has made use of the NASA/IPAC Extragalactic Database (NED) which is operated by the Jet Propulsion Laboratory, California Institute of Technology, under contract with the National Aeronautics and Space Administration.
Some of the data presented in this paper were obtained from the Multimission Archive at the Space Telescope Science Institute (MAST). Support for MAST for non-HST data is provided by the NASA Office of Space Science via grant NAG5-7584 and by other grants and contract.

{\it Facilities:} \facility{Bok, CTIO:0.9m, GALEX, Spitzer, VATT}.

\clearpage
\clearpage
\clearpage

\begin{deluxetable}{clcrcccccccccc}
\tabletypesize{\scriptsize}
\rotate
\tablecolumns{14}
\tablewidth{0pc}
\tablecaption{11HUGS H$\alpha$ and UV SFRs}

\tablehead{

\colhead{\#}            &
\colhead{Galaxy name}   &
\colhead{RA}            &
\colhead{DEC}           &
\colhead{$E(B-V)$}      &
\colhead{D}             &
\colhead{method}        &
\colhead{$M_B$}         &
\colhead{T}             &
\colhead{A(H$\alpha$)}  &
\colhead{A(FUV)}        &
\colhead{code}          &
\colhead{SFR(UV)}       &
\colhead{SFR(H$\alpha$)/SFR(UV)}
\\

\colhead{}              &  
\colhead{}              &
\colhead{[J2000]}       &
\colhead{[J2000]}       &
\colhead{[mag]}         &
\colhead{[Mpc]}         &
\colhead{}              &
\colhead{[mag]}         &
\colhead{}              &
\colhead{[mag]}         &
\colhead{[mag]}         &
\colhead{}              &
\colhead{LOG [M$_{\odot}$~yr$^{-1}$]}       &
\colhead{}  

\\

\colhead{(1) }           &  
\colhead{(2) }           &
\colhead{(3) }           &
\colhead{(4) }           &
\colhead{(5) }           &
\colhead{(6) }           &
\colhead{(7) }           &
\colhead{(8) }           &
\colhead{(9) }           &
\colhead{(10) }          &
\colhead{(11) }          &
\colhead{(12) }          &
\colhead{(13) }          &
\colhead{(14) }          
}

\startdata
    1 & WLM & 000158.1 & -152739 & 0.04 &0.92 & trgb & -13.94 &10 &0.10 &\nodata &\nodata &     -2.21 &      0.31 \\ 
     2 & ESO409-IG015 & 000531.8 & -280553 & 0.02 &10.4 & v(flow) & -15.0 & 6 &0.14 &\nodata &\nodata &     -1.45 &      1.48 \\ 
     3 & NGC24 & 000956.7 & -245744 & 0.02 & 8.1 & v(flow) & -17.4 & 5 &0.41 &0.74 &\nodata &     -0.70 &      0.56 \\ 
     4 & NGC45 & 001404.0 & -231055 & 0.02 & 7.1 & v(flow) & -18.01 & 8 &0.50 &0.30 &\nodata &     -0.40 &      0.99 \\ 
     5 & NGC55 & 001454.0 & -391149 & 0.01 &2.17 & trgb & -18.32 & 9 &0.55 &0.86 &\nodata &     -0.26 &      0.86 \\ 

\enddata 
\tablecomments{
Col.(1): Running index number in this table. 
Col.(2): Galaxy name. 
Col.(3-4): J2000 R.A. and Dec. from NED. 
Col.(5): Milky Way reddening based on the maps of Schlegel et al. (1998). 
Col.(6): Distance to the galaxy in Mpc. 
Col.(7): Method of distance determination as compiled, described, and referenced in Paper I. $H_o=$75 km s$^{-1}$ Mpc$^{-1}$ is assumed for distances estimated from Local Group flow corrected recessional velocities. 
Col.(8): Absolute $B$-band magnitude based on the photometry compiled in Paper I, and the distances and Milky Way reddenings given the previous columns. 
Col.(9): RC3 Morphological T-type, compiled as described in Paper I.  
Col.(10): H$\alpha$ nebular attentuation in magnitudes calculated as described in Section~\label{sec:dustcorr}. 
Col.(11): FUV attentuation in magnitudes based on the TIR/FUV ratio as described in Section~\label{sec:dustcorr}.  
Col.(12): Source of the H$\alpha$/H$\beta$ Balmer line ratio used to calculate the H$\alpha$ nebular attentuation. ``mk" refers to the integrated spectral atlas of Moustakas and Kennicutt (2006), and ``nfgs" refers to the integrated spectra of the Nearby Field Galaxy Survey (Jansen et al. 2000).  A dash preceding the reference ("-mk" or "-nfgs") indicates that H$\alpha$/H$\beta$ is less than the expected Case B ratio of 2.86.  This results in a negative extinction correction, so in these cases the value of A(H$\alpha$) given is estimated from $M_B$, as are those where spectral line ratio measurements are not available from one of these sources. 
Col.(13): SFR, in units of log[M$_{\odot}$~yr$^{-1}$], based on the conversion of K98, the integrated GALEX FUV magnitude reported in Paper II and the A(FUV) given in this table.  If no value for A(FUV) is listed, the attenuation is estimated from 1.8*A(H$\alpha$).  
Col.(14): Ratio of SFRs calculated from H$\alpha$ and the FUV.  The H$\alpha$ SFRs are based on the conversion of K98, line fluxes and [NII]H$\alpha$ corrections given in Paper I, and the A(H$\alpha$) listed in this table.
}
\end{deluxetable}


\clearpage
\begin{deluxetable}{cccr}
\tabletypesize{\scriptsize}
\tablecolumns{3}
\tablewidth{0pc}
\tablecaption{H$\alpha$-to-UV SFR Ratios\label{tab:binnedavgs}}

\tablehead{
\multicolumn{3}{c}{\it Observed}  \\ 
\hline
\colhead{LOG SFR(H$\alpha$)}   &
\colhead{LOG $\frac{\mbox{SFR(H$\alpha$)}}{\mbox{SFR(UV)}}$}&
\colhead{1$\sigma$-scatter} &
\colhead{$N_{gal}$}
}

\startdata
 0.25 &  0.20 &  0.37 &   11 \\ 
-0.25 &  0.17 &  0.30 &   21 \\ 
-0.75 &  0.07 &  0.26 &   39 \\ 
-1.25 & -0.02 &  0.25 &   66 \\ 
-1.75 & -0.10 &  0.22 &   55 \\ 
-2.25 & -0.23 &  0.22 &   61 \\ 
-2.75 & -0.46 &  0.26 &   33 \\ 
-3.50 & -0.49 &  0.58 &   20 \\ 
-4.50 & -1.29 &  0.57 &    6 \\ 

\hline
$M_B$ & LOG $\frac{\mbox{SFR(H$\alpha$)}}{\mbox{SFR(UV)}}$& 1$\sigma$-scatter&$N_{gal}$\\
\hline
-20.5 &  0.18 &  0.23 &   14 \\ 
-19.5 &  0.20 &  0.39 &   12 \\ 
-18.5 &  0.08 &  0.34 &   27 \\ 
-17.5 & -0.01 &  0.31 &   36 \\ 
-16.5 &  0.00 &  0.21 &   35 \\ 
-15.5 & -0.22 &  0.37 &   58 \\ 
-14.5 & -0.26 &  0.29 &   61 \\ 
-13.5 & -0.31 &  0.30 &   41 \\ 
-12.5 & -0.41 &  0.66 &   21 \\ 
-11.5 & -0.48 &  0.61 &    7 \\ 

\hline
\multicolumn{3}{c}{\it Dust Corrected}\\
\hline
LOG SFR(H$\alpha$) & LOG $\frac{\mbox{SFR(H$\alpha$)}}{\mbox{SFR(UV)}}$& 1$\sigma$-scatter&$N_{gal}$\\
\hline
 0.50 & -0.14 &  0.29 &   20 \\ 
-0.25 & -0.09 &  0.20 &   26 \\ 
-0.75 & -0.13 &  0.22 &   44 \\ 
-1.25 & -0.13 &  0.22 &   51 \\ 
-1.75 & -0.18 &  0.17 &   58 \\ 
-2.25 & -0.27 &  0.22 &   56 \\ 
-2.75 & -0.51 &  0.25 &   31 \\ 
-3.50 & -0.55 &  0.57 &   21 \\ 
-4.50 & -1.43 &  0.59 &    5 \\ 

\hline
$M_B$ & LOG $\frac{\mbox{SFR(H$\alpha$)}}{\mbox{SFR(UV)}}$& 1$\sigma$-scatter&$N_{gal}$\\
\hline
-20.5 & -0.21 &  0.23 &   14 \\ 
-19.5 & -0.10 &  0.36 &   12 \\ 
-18.5 & -0.10 &  0.23 &   27 \\ 
-17.5 & -0.16 &  0.17 &   36 \\ 
-16.5 & -0.12 &  0.18 &   35 \\ 
-15.5 & -0.27 &  0.36 &   58 \\ 
-14.5 & -0.30 &  0.28 &   61 \\ 
-13.5 & -0.34 &  0.30 &   41 \\ 
-12.5 & -0.45 &  0.67 &   21 \\ 
-11.5 & -0.51 &  0.61 &    7 \\ 

\enddata

\end{deluxetable}

\clearpage

\end{document}